\documentclass[a4paper,11pt]{article}
\pdfoutput=1
\usepackage{jheppub}

\usepackage{dsfont}
\usepackage{graphicx}
\usepackage{empheq}
\usepackage{caption}
\usepackage{subcaption}
\usepackage{todonotes}
\usepackage{mathtools}

\usepackage{tikz}
\usetikzlibrary{arrows,shapes}
\usetikzlibrary{trees}
\usetikzlibrary{patterns}
\usetikzlibrary{matrix,arrows} 				
\usetikzlibrary{positioning}				
\usetikzlibrary{calc,through}				
\usetikzlibrary{decorations.pathreplacing}  
\usepackage{pgffor}							

\usetikzlibrary{decorations.pathmorphing}	
\usetikzlibrary{decorations.markings}
\tikzset{
    vector/.style={decorate, decoration={snake}, draw},
    graviton/.style={decorate, double, decoration={snake}, draw},
	provector/.style={decorate, decoration={snake,amplitude=2.5pt}, draw},
	antivector/.style={decorate, decoration={snake,amplitude=-2.5pt}, draw},
        smallvector/.style={decorate, decoration={snake,amplitude=1.5pt,post length=0.5mm}, draw},
    fermion/.style={draw=black, postaction={decorate},
        decoration={markings,mark=at position .55 with {\arrow[draw=black]{>}}}},
    fermionbar/.style={draw=black, postaction={decorate},
        decoration={markings,mark=at position .55 with {\arrow[draw=black]{<}}}},
    fermionnoarrow/.style={draw=black},
    gluon/.style={decorate, draw=black,
        decoration={coil,amplitude=4pt, segment length=5pt}},
    scalar/.style={dashed,draw=black, postaction={decorate},
        decoration={markings,mark=at position .55 with {\arrow[draw=black]{>}}}},
    scalarbar/.style={dashed,draw=black, postaction={decorate},
        decoration={markings,mark=at position .55 with {\arrow[draw=black]{<}}}},
    scalarnoarrow/.style={dashed,draw=black},
    electron/.style={draw=black, postaction={decorate},
        decoration={markings,mark=at position .55 with {\arrow[draw=black]{>}}}},
    bigvector/.style={decorate, decoration={snake,amplitude=4pt}, draw},
    arrow/.style={draw=black, postaction={decorate},
        decoration={markings,mark=at position 1 with {\arrow[draw=black]{>}}}},
}

\tikzstyle{block} = [draw, rectangle, 
    minimum height=3em, minimum width=6earticlem]

\renewcommand{\Im}{\operatorname{Im}}
\usepackage{xcolor}
\renewcommand{\imath}{\mathrm{i}}
\newcommand{\vect}{\boldsymbol}

\let\originalleft\left
\let\originalright\right
\renewcommand{\left}{\mathopen{}\mathclose\bgroup\originalleft}
\renewcommand{\right}{\aftergroup\egroup\originalright}

\begin{document}

\title{Absorptive Effects and Classical Black Hole Scattering}
\author[a]{Callum R. T. Jones,}
\emailAdd{cjones@physics.ucla.edu}
\author[a]{Michael S. Ruf,}
\emailAdd{mruf@physics.ucla.edu}
\affiliation[a]{
Mani L. Bhaumik Institute for Theoretical Physics, Department of Physics and Astronomy, University of California Los Angeles, Los Angeles, CA 90095, USA
}

\abstract{We describe an approach to incorporating the physical effects of the absorption of energy by the event horizon of black holes in the scattering amplitudes based post-Minkowskian, point-particle effective description. Absorptive dynamics are incorporated in a model-independent way by coupling the usual point-particle description to an invisible sector of gapless internal degrees-of-freedom. The leading order dynamics of this sector are encoded in the low-energy expansion of a spectral density function obtained by matching an absorption cross section in the ultraviolet description. This information is then recycled using the scattering amplitudes based Kosower-Maybee-O'Connell in-in formalism to calculate the leading absorptive contribution to the impulse and change in rest mass of a Schwarzschild black hole scattering with a second compact body sourcing a massless scalar, electromagnetic or gravitational field. The results obtained are in complete agreement with previous worldline Schwinger-Keldysh calculations and provide an alternative on-shell scattering amplitudes approach to incorporating horizon absorption effects in the gravitational two-body problem.
}

\maketitle
\flushbottom


\section{Introduction}
\label{sec:intro}

The detection of gravitational waves by the LIGO and Virgo collaborations has inaugurated a new era of gravitational wave astrophysics \cite{LIGOScientific:2016aoc,LIGOScientific:2017vwq}. Beyond the celebrated initial discovery, the next generation of ground- and space-based gravitational wave detectors \cite{Punturo:2010zz,2017arXiv170200786A,Reitze:2019iox} are expected to increase sensitivity by two orders-of-magnitude, beginning an era of \textit{precision} gravitational wave science. The success of this experimental program will require commensurate advances in theoretical waveform predictions. A powerful approach to generating these predictions is given by the effective one-body (EOB) resummation \cite{Buonanno:1998gg,Buonanno:2000ef}, taking as input information from a variety of different physical regimes including non-perturbative numerical relativity simulation \cite{Pretorius:2005gq,baumgarte_shapiro_2010,Damour:2014afa,Hopper:2022rwo}, self-force expansion \cite{Mino:1996nk,Quinn:1996am,Poisson:2011nh,Barack:2018yvs} and perturbative weak-field calculations based on the post-Newtonian (PN) \cite{Einstein:1938yz,Ohta:1973je} and post-Minkowskian (PM) expansions \cite{Bertotti:1956pxu,Kerr:1959zlt,Bertotti:1960wuq,Westpfahl:1979gu,Portilla:1980uz,Bel:1981be}.

Recent years have seen tremendous advances in these weak-field perturbative approaches. In particular, new and powerful calculational methods based on relativistic scattering amplitudes \cite{Cheung:2018wkq,Bern:2019nnu,Bern:2019crd,DiVecchia:2020ymx,DiVecchia:2021bdo,Bjerrum-Bohr:2021din,Brandhuber:2021eyq,Herrmann:2021lqe,DiVecchia:2021ndb,Heissenberg:2021tzo,Bern:2021dqo,Bern:2021yeh,Jones:2022aji} and worldline effective field theories (EFT) \cite{Kalin:2020mvi,Kalin:2020fhe,Dlapa:2021npj,Mogull:2020sak,Jakobsen:2021smu,Jakobsen:2022psy,Loebbert:2020aos,Dlapa:2021vgp,Dlapa:2022lmu,Kalin:2022hph}, have allowed for rapid progress in understanding the PM scattering regime of the gravitational two-body problem. These approaches share the common strategy of initially framing the problem in \textit{quantum mechanical} language, and then using the eventual \textit{classical} or $\hbar \rightarrow 0$ limit as an additional expansion leading to vast simplifications in the required calculations. This quantum-first perspective leads to a natural synthesis with techniques developed in modern high-energy physics including the systematic use of EFTs \cite{Goldberger:2004jt,Neill:2013wsa,Cheung:2018wkq,Damgaard:2019lfh,Goldberger:2022rqf}, unitarity-based methods \cite{Bern:1994zx,Bern:1994cg,Bern:1995db,Bern:1997sc,Britto:2004nc}, double-copy \cite{Kawai:1985xq,Bern:2008qj,Bern:2010ue,Bern:2019prr} and methods for advanced multi-loop integration \cite{Chetyrkin:1981qh,Laporta:2000dsw,Kotikov:1990kg,Bern:1993kr,Remiddi:1997ny,Gehrmann:1999as}. Further details can be found in the review \cite{Buonanno:2022pgc}. 

Physically, the PM scattering regime is an expansion based on a separation of scales between the intrinsic size of the compact bodies $R$ ($=2G_\mathrm{N} M$ for a Schwarzschild black hole) and the impact parameter $b$. In the limit $R\ll b$ the problem admits an effective description as the scattering of point-particles interacting by exchanging gravitons (and possibly other massless force-mediating particles). At low orders in this long-distance expansion it is sufficient to calculate scattering observables from a model in which the compact bodies are mathematically represented as structureless elementary particles minimally coupled to the mediator fields. At higher-orders, finite size effects will begin to contribute; some of these effects can be captured by modifying the effective description with non-minimal couplings (permanent multipole moments or tidal Love numbers) \cite{Cheung:2020sdj,Bern:2020uwk}, but some cannot. As emphasized long-ago \cite{Goldberger:2005cd}, even in the long-distance regime, classical macroscopic bodies, including black holes and neutron stars, differ qualitatively from elementary particles due to the existence of \textit{gapless} internal degrees-of-freedom \cite{PhysRev.83.34,Goldberger:2022rqf}.

As a motivating illustration, consider a lump of some material modelled as an Einstein solid, where the internal degrees-of-freedom are approximated as a large number of independent quantum harmonic oscillators \cite{https://doi.org/10.1002/andp.19063270110}. The quantum mechanical spectrum of this model is discrete with level spacing of $\mathcal{O}\left(\hbar\right)$. In the classical limit $\hbar \rightarrow 0$, this model has an effectively continuous spectrum of excited states extending all the way to the ground state threshold without a gap. Even though a precise microscopic description of a Schwarzschild black hole is not known, we expect that it shares these qualitative features; from explicit black hole perturbation theory calculations it is known that a Schwarzschild black hole can absorb radiation of arbitrarily low frequency \cite{Starobinskil:1974nkd,Page:1976df}, necessitating the existence of (classically) gapless excited states. 

In a two-body scattering event, no matter how small the momentum transfer between the bodies, there will necessarily exist near-threshold excited states that can go on-shell and therefore cannot be integrated out of the point-particle effective theory. Since we may have neither direct experimental access to these states, nor a precise microscopic theoretical description, our approach to incorporating their physical effects is to regard them as constituting an \textit{invisible sector} into which energy and other quantum numbers may be absorbed. As our choice of language suggests, this problem shares many formal similarities with model-independent approaches to describing interactions between the Standard Model and a hidden sector \cite{Contino:2020tix,Arina:2021nqi,Klose:2022vro}. From this perspective, the problem of scattering black holes and other compact macroscopic bodies is to be treated as the evolution of an open quantum system. 

In this paper we develop a scattering amplitudes based formalism for incorporating the physical effects of these near-threshold excited states, closely modelled on the worldline formalism described in \cite{Goldberger:2005cd,Porto:2007qi,Goldberger:2019sya,Goldberger:2020wbx,Goldberger:2020fot}. At leading PM order the effects of absorption are parametrized by the low-energy expansion of a spectral density function obtained by explicitly matching with an absorption cross section calculated in black hole perturbation theory \cite{Starobinskil:1974nkd,Page:1976df}. We then, in an EFT sense, recycle this information to calculate distinct low-energy observables. In particular, in this paper we calculate the leading PM absorptive contributions to the impulse of a Schwarzschild black hole scattering with a second compact body sourcing a massless scalar (\ref{massshiftscalar}), electromagnetic (\ref{massshiftphoton}) or gravitational field (\ref{massshiftgraviton}). The contribution to the impulse from graviton absorption agrees perfectly with previous worldline calculations \cite{Goldberger:2020wbx}; as far as we are aware the expressions for the impulse due to scalar and photon absorption are new. 

Previous theoretical studies of horizon absorption effects include \cite{Poisson:1994yf,Alvi:2001mx,Poisson:2004cw,Hughes:2001jr,Nagar:2011aa,Bernuzzi:2012ku,Taracchini:2013wfa,Goldberger:2005cd,Porto:2007qi,Endlich:2016jgc,Goldberger:2019sya,Goldberger:2020wbx,Goldberger:2020fot,Saketh:2022xjb,Aoude:2023fdm,Tagoshi:1997jy,Chatziioannou:2016kem,Chia:2020yla,Saketh:2023bul,Bautista:2023szu}. The problem of calculating the absorptive contribution to the impulse in two-body scattering was approached in \cite{Goldberger:2020wbx} using off-shell worldline based methods, where in-in observables are naturally calculated using the Schwinger-Keldysh or closed-time-path formalism \cite{Schwinger:1960qe,Keldysh:1964ud,Galley:2009px}. In this paper we approach this problem from an on-shell scattering amplitudes perspective, where the natural in-in formalism is given by Kosower, Maybee and O'Connell (KMOC) \cite{Kosower:2018adc}.  As emphasized recently \cite{Damgaard:2023vnx} these formalisms are closely related, but organize the calculation in very different ways. Even though we are calculating an in-in observable, KMOC takes ordinary in-out scattering amplitudes as building blocks, allowing us to retain some of the power and simplicity of modern amplitudes methods and has a natural synthesis with the previously mentioned scattering amplitudes based approaches to the gravitational two-body problem.

This paper is organized as follows. In Section \ref{sec:open} we introduce a general framework for incorporating horizon absorption by coupling the point-particle effective description to an invisible sector. In Section \ref{subsec:crosssection} we demonstrate that the leading-order low-energy dynamics of this sector are encoded in the expansion of a spectral density function and obtain this information by explicitly matching with an absorption cross section. The absorptive contribution to the impulse of a black hole during a two-body scattering event is then obtained in Section \ref{sec:impulse}. We explain how absorption of energy by the invisible sector degrees-of-freedom can be naturally incorporated in the KMOC formalism in Section \ref{subsec:KMOC}. We then describe in Sections \ref{subsec:Love} and \ref{subsec:light}, how the ``heavy" and ``light" invisible sector degrees of freedom manifest as Love number type contact contributions and absorptive effects respectively. In \ref{subsec:mass} the necessary box diagrams are constructed using unitarity-based methods and the resulting impulses and mass-shifts calculated for the absorption of scalars, photons and gravitons. Finally, in Section \ref{subsec:neutron} we generalize the discussion to generic compact bodies including neutron stars, parametrizing the leading-order impulse and mass-shifts in terms of dissipation numbers.

\section{Classical Black Holes as Open Quantum Systems}
\label{sec:open}

\subsection{Visible and Invisible Degrees-of-Freedom}
\label{subsec:invisible}

We will consider the scattering of a generic compact body $\phi_1$ with a Schwarzschild black hole $\phi_2$ with masses $m_1$ and $m_2$ respectively. In addition to the gravitational field, body-1 may source an electromagnetic field $A_\mu$ with electric charge $Q_e$ and a massless scalar field $\psi$ with scalar charge $Q_s$. These degrees-of-freedom constitute what we will call the \textit{visible sector}, and are described by an action
\begin{align}
    \label{Svis}
    S_{\text{vis}} &= \int \text{d}^4 x \sqrt{-g}\left[-\frac{2}{\kappa^2}R-\frac{1}{4}F_{\mu\nu}^2+\frac{1}{2}\left(\nabla_\mu \psi\right)^2+|D_\mu \phi_1|^2 -m_1^2 |\phi_1|^2 \right.\nonumber\\
    &\hspace{27mm}\left.+ \frac{1}{2}\left(\nabla_\mu \phi_2\right)^2-\frac{1}{2}m_2^2 \phi_2^2 +Q_s \psi|\phi_1|^2\right] + S_{\text{HD}} + S_{\text{GF}},
\end{align}
where $\kappa \coloneqq\sqrt{32\pi G_\mathrm{N}}$, $D_\mu \coloneqq\nabla_\mu +\imath Q_e A_\mu$ and $F_{\mu\nu} \coloneqq\nabla_\mu A_\nu - \nabla_\nu A_\mu$; $S_{\text{GF}}$ are gauge fixing terms and $S_{\text{HD}}$ denotes non-minimal (tidal Love number) interactions \cite{Cheung:2020sdj,Bern:2020uwk}. As discussed in Section \ref{sec:intro}, at low orders in the PM expansion this action captures the relevant physics of scattering compact bodies at long distances. At some order in $G_\mathrm{N}$ (to be determined), the physical effects of horizon absorption become important; physical transitions between the black hole ground state and an excited state become possible and these cannot be described by an action of the form (\ref{Svis}).

The Hilbert space in this problem takes a factored form $\mathcal{H}_{\text{vis}}\otimes \mathcal{H}_{\text{inv}}$, corresponding to visible states (black hole translation/rotation modes and soft radiation) and invisible states (gapless internal degrees-of-freedom). There are many different approaches to describing the dynamics of this system, we could explicitly trace over $\mathcal{H}_{\text{inv}}$ and describe the scattering problem as the non-unitary time evolution of a reduced density matrix in $\mathcal{H}_{\text{vis}}$. To retain some of the simplicity of unitary time evolution we instead choose to work on the full Hilbert space, calculating in-in observables where the initial state is in the visible sector. The invisible states, that we will collectively denote as $X$, appear only as internal states to be summed over. 

Our formalism for incorporating these invisible sector states is modelled on the worldline formalism described in \cite{Goldberger:2005cd,Porto:2007qi,Goldberger:2019sya,Goldberger:2020wbx,Goldberger:2020fot}. The complete action for the system consists of the the visible sector (\ref{Svis}), the unknown (possibly strongly coupled) self-interactions of the invisible $X$-states denoted $S_{\text{inv}}$ and \textit{portal} couplings between the visible and invisible sectors. For the latter, the $X$-states are encoded in the form of abstract,  composite, local operators $\mathcal{O}_i(x)$; for which we assume the usual properties of Poincar\'{e} invariance, gauge/diffeomorphism invariance and locality. Without loss of generality we exclude quadratic portal couplings e.g. $\phi_2 \mathcal{O}$; such interactions can always be removed by redefining the visible sector field basis. To leading order in $G_\mathrm{N}$ the portal couplings are given by 
\begin{align}
    \label{Svisinv}
    S_{\text{portal}} &= \int \text{d}^4 x\sqrt{-g} \left[ \kappa\phi_2 \psi \mathcal{O}_0 + \kappa\phi_2 F_{\mu\nu}\mathcal{O}_1^{\mu\nu} + \phi_2 C_{\mu\nu\rho\sigma}\mathcal{O}_2^{\mu\nu\rho\sigma}  +\cdots\right],
\end{align}
where $C_{\mu\nu\rho\sigma}$ is the Weyl tensor. The idea is then to calculate observables by evaluating the path integral for the visible sector, treating the portal couplings perturbatively in $G_\mathrm{N}$ and leaving the invisible sector path integral unevaluated. As an illustrative example, consider the calculation of the off-shell Green's function 
\begin{equation}
    \label{4point}
    \langle \phi_2(x_1) \psi(x_2) \phi_2(x_1') \psi(x_2')\rangle = \int [\mathcal{D}\phi_2][\mathcal{D}\psi][\mathcal{D}X]\phi_2 (x_1) \psi(x_2) \phi_2(x_1') \psi(x_2')e^{\imath S_{\text{vis}} + \imath S_{\text{inv}}+ \imath S_{\text{portal}}}.
\end{equation}
Diagrammatically the leading connected contribution from the $X$-states takes the form: \\
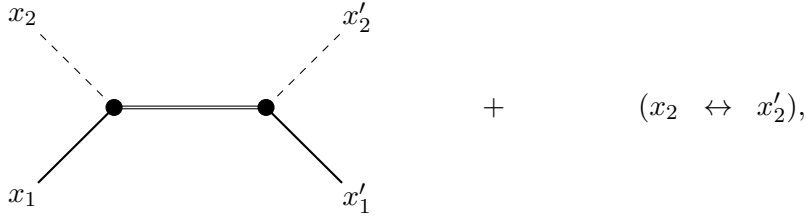
\begin{center}
    \begin{tikzpicture}
        \draw [-,thick] (-1,-1)--(0,0);
        \draw [-, double] (0,0)--(2,0);
        \filldraw (0,0) circle (3pt);
        \filldraw (2,0) circle (3pt);
        \draw [-,thick] (2,0)--(3,-1);
        \draw [dashed] (0,0)--(-1,1);
        \draw [dashed] (2,0)--(3,1);
        \node at (-1.2,1.2) {$x_2$};
        \node at (-1.2,-1.2) {$x_1$};
        \node at (3.2,1.2) {$x_2'$};
        \node at (3.2,-1.2) {$x_1'$};
        \node at (5,0) {$+$};
        \node at (8,0) {$\left(x_2 \;\;\leftrightarrow\;\; x_2'\right),$};
    \end{tikzpicture}
\end{center}
where the internal doubled line corresponds to the (non-perturbative) invisible sector 2-point function
\begin{equation}
    \langle \mathcal{O}_0(x)\mathcal{O}_0(0)\rangle_X \coloneqq\int [\mathcal{D}X] \mathcal{O}_0(x)\mathcal{O}_0(0) e^{\imath S_{\text{inv}}}.
\end{equation}
It is natural to rewrite this in \textit{K\"{a}ll\'{e}n-Lehmann} (KL) form 
\begin{equation}
    \label{O1O1}
    \langle \mathcal{O}_0(x)\mathcal{O}_0(0)\rangle_X = \int \hat{\text{d}}^D k \; e^{\imath k\cdot x} \int_{m_2^2}^\infty \text{d}\mu^2 \frac{\imath\rho_0(\mu^2)}{k^2-\mu^2+\imath 0},
\end{equation}
where the unknown invisible sector physics is contained in the \textit{spectral density} $\rho_0(\mu^2)$. In this form, momentum space calculations involving internal $X$-states can be performed using a small modification of standard Feynman rules, summarized in Appendix \ref{app:Feynman}, incorporating weighted integration over the spectral parameter $\mu^2$. 

\subsection{UV Matching: Absorption Cross Section}
\label{subsec:crosssection}

To determine the spectral density we match an \textit{absorption cross section} calculated in both the UV (full General Relativity) and the IR (the point-particle effective description). To leading order in $\omega \rightarrow 0$, where $\omega$ is the energy of the absorbed radiation, the necessary Schwarzschild black hole absorption cross sections were calculated long ago \cite{Starobinskil:1974nkd,Page:1976df}
\begin{equation}
\label{sigmaUV}
    \sigma_{\text{abs}}\left(\omega\right) \sim 
    \begin{cases}
        16\pi G_\mathrm{N}^2 m_2^2 & \text{for scalars,}\\
        \frac{64\pi}{3}G_\mathrm{N}^4 m_2^4 \omega^2 & \text{for photons,}\\
        \frac{256\pi}{45}G_\mathrm{N}^6 m_2^6 \omega^4 & \text{for gravitons.}
    \end{cases}
\end{equation}
For the photon and graviton cases the above cross sections correspond to an incoming polarized state with helicity $\pm 1$ and $\pm 2$ respectively, the explicit expression is independent of the polarization state. In the point-particle effective description, the corresponding absorption cross sections are calculated perturbatively; the general procedure is illustrated first for the absorption of a massless scalar, and then the particular complications of photon and graviton absorption described separately.

\subsubsection*{Scalar Absorption Cross Section}

For scalar absorption we first calculate the Compton amplitude 
\begin{equation}
  \mathcal{M}\left(\phi_2\left(-p_2\right) \psi\left(-k_1\right) \rightarrow \phi_2\left(p_3\right) \psi\left(k_4\right)\right),
\end{equation}
where the external momenta are lablelled in the all-outgoing convention. Similar to the off-shell Green's function (\ref{4point}) the contribution of the visible sector states is calculated perturbatively, while the invisible $X$-states contribute through the KL ``propagator" (\ref{O1O1}). In the forward limit kinematics
\begin{equation}
    \label{forwardkin}
    -{p}_2^\mu = p_3^\mu = \left(m_2,0,0,0\right),\hspace{10mm} -{k}_1^\mu = k_4^\mu = \left(\omega,0,0,\omega\right),
\end{equation}
the optical theorem relates the imaginary part of the Compton amplitude to the total cross section
\begin{equation}
    \sigma_{\text{total}}\left(\omega\right) = \frac{\text{Im}\mathcal{M}^{\text{forward}}\left(\omega\right)}{2m_2 \omega}.
\end{equation}
At leading order the absorptive part of the cross section can be disentangled from the elastic part by considering the contribution of the diagram \\
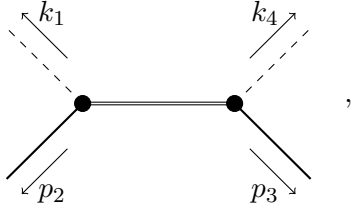
\begin{center}
    \begin{tikzpicture}
        \draw [-,thick] (-1,-1)--(0,0);
        \draw [-, double] (0,0)--(2,0);
        \filldraw (0,0) circle (3pt);
        \filldraw (2,0) circle (3pt);
        \draw [-,thick] (2,0)--(3,-1);
        \draw [dashed] (0,0)--(-1,1);
        \draw [dashed] (2,0)--(3,1);
        \draw[<-] (-0.8,1.2)--(-0.2,0.6);
        \draw[<-] (-0.8,-1.2)--(-0.2,-0.6);
        \draw[->] (2.2,0.6)--(2.8,1.2);
        \draw[->] (2.2,-0.6)--(2.8,-1.2);
        \node at (-0.4,1.2) {$k_1$};
        \node at (-0.4,-1.2) {$p_2$};
        \node at (2.4,1.2) {$k_4$};
        \node at (2.4,-1.2) {$p_3$};
        \node at (3.5,0) {,};
    \end{tikzpicture}
\end{center}
in the forward kinematics (\ref{forwardkin}) this corresponds to
\begin{equation}
    \mathcal{M}^{(\psi)}\biggr\vert_{\text{forward}} = -\kappa^2\int_{m_2^2}^\infty \text{d}\mu^2 \frac{\rho_0(\mu^2)}{m_2^2 +2m_2 \omega-\mu^2+\imath 0}.
\end{equation}
Using the distributional identity
\begin{equation}
    \text{Im}\left(\frac{1}{x+\imath 0}\right) = -\pi \delta(x),
\end{equation}
we obtain the imaginary part
\begin{equation}
    \text{Im}\mathcal{M}^{(\psi)}\biggr\vert_{\text{forward}} = \pi\kappa^2 \int_{m_2^2}^\infty \text{d}\mu^2 \rho_0(\mu^2)\delta\left(m_2^2 +2m_2 \omega-\mu^2\right) = \pi \kappa^2 \rho_0\left(m_2^2+2m_2\omega\right),
\end{equation}
and therefore the absorption cross section
\begin{equation}
    \sigma_{\text{abs}}^{(\psi)}\left(\omega\right) = \frac{\pi \kappa^2\rho_0\left(m_2^2+2m_2\omega\right)}{2m_2 \omega}.
\end{equation}
Matching this with the UV cross section (\ref{sigmaUV}) gives the near-threshold asymptotic expansion of the spectral density
\begin{equation}
    \label{rho0}
   \rho_0\left(\mu^2\right) \sim \frac{G_\mathrm{N} m_2^2 }{2\pi} \left(\mu^2-m_2^2\right), \hspace{5mm}\text{as} \hspace{5mm} \mu^2 \rightarrow m_2^2.
\end{equation}

\subsubsection*{Photon Absorption Cross Section}

For photon absorption, the composite operator $\mathcal{O}_1^{\mu\nu}$ that appears in (\ref{Svisinv}) transforms non-trivially under the Lorentz group and therefore has a more complicated KL form
\begin{equation}
    \langle \mathcal{O}_1^{\mu\nu}(x) \mathcal{O}_1^{\rho\sigma}(0)\rangle_X = \int \hat{\text{d}}^D k \; e^{ \imath k\cdot x} \int_{m_2^2}^\infty \text{d}\mu^2 \frac{\imath\Pi_1^{\mu\nu\rho\sigma}\left(\mu^2,k\right)}{k^2-\mu^2+\imath 0}.
\end{equation}
In general the projector $\Pi_1$ is constructed from the available Lorentz tensors $k^\mu$, $\eta^{\mu\nu}$ and $\epsilon^{\mu\nu\rho\sigma}$ and constrained by the symmetry of the operators. Without loss of generality we can assume that $\mathcal{O}_1^{\mu\nu} = -\mathcal{O}_1^{\nu\mu}$ and we will impose that the 2-point function is parity invariant. Together with the exchange symmetry
\begin{equation}
    \Pi_1^{\mu\nu\rho\sigma}\left(\mu^2,k\right) = \Pi_1^{\rho\sigma\mu\nu}\left(\mu^2,k\right),
\end{equation}
we find there are two tensor structures compatible with these assumptions 
\begin{align}
    \Pi_1^{\mu\nu\rho\sigma}\left(\mu^2,k\right) &= \rho_{1}^{(1)}\left(\mu^2\right)\left[\eta^{\mu\rho}\eta^{\nu\sigma}-\eta^{\mu\sigma}\eta^{\nu\rho}\right]\nonumber\\
    &\hspace{5mm}+\rho_{1}^{(2)}\left(\mu^2\right)\left[\frac{k^\nu k^\rho \eta^{\mu\sigma}}{k^2}+\frac{k^\mu k^\sigma \eta^{\nu\rho}}{k^2}-\frac{k^\mu k^\rho \eta^{\nu\sigma}}{k^2}-\frac{k^\nu k^\sigma \eta^{\mu\rho}}{k^2}\right].
\end{align}
As discussed further in Section \ref{subsec:neutron} for a generic macroscopic body capable of absorbing electromagnetic radiation (e.g. a neutron star or a dielectric sphere with complex permittivity) there are two independent spectral functions that must be matched to the UV description. As described in \cite{Endlich:2016jgc,Aoude:2023fdm} the independent contributions can be obtained by separately matching contributions to the absorption cross section from incoming spinning partial waves. For Schwarzschild black holes in $d=4$ this is actually unnecessary due to the presence of a hidden symmetry, the \textit{electromagnetic self-duality} of the Einstein-Maxwell equations \cite{Deser:1976iy}. Since this is a symmetry of the UV model we impose that it is also a symmetry of the effective point-particle description. Imposing duality invariance of the 2-point function 
\begin{equation} 
\label{emduality}
    \Pi_1^{\mu\nu\rho\sigma}\left(\mu^2,k\right) = \frac{1}{4}{\epsilon^{\mu\nu}}_{\alpha\beta}{\epsilon^{\rho\sigma}}_{\gamma\delta}\Pi_1^{\alpha\beta\gamma\delta}\left(\mu^2,k\right),
\end{equation}
gives a relation between the spectral functions $\rho_1^{(1)}$ and $\rho_1^{(2)}$. The unique duality invariant tensor structure is found to be
\begin{equation}
    \Pi_1^{\mu\nu\rho\sigma}\left(\mu^2,k\right) = \rho_1\left(\mu^2\right)\; \hat{\Pi}_1^{\mu\nu\rho\sigma}\left(k\right),
\end{equation}
where
\begin{equation}
    \label{hatPi1}
    \hat{\Pi}_1^{\mu\nu\rho\sigma}\left(k\right) \coloneqq\frac{1}{2}\left(\eta^{\mu\rho}\eta^{\nu\sigma}-\eta^{\mu\sigma}\eta^{\nu\rho}\right)+\frac{1}{k^2}\left(k^\nu k^\rho \eta^{\mu\sigma}+k^\mu k^\sigma \eta^{\nu\rho}-k^\mu k^\rho \eta^{\nu\sigma}-k^\nu k^\sigma \eta^{\mu\rho}\right).
\end{equation}
In the context of modern unitarity-based methods \cite{Bern:1994zx,Bern:1994cg,Bern:1995db,Bern:1997sc,Britto:2004nc}, the natural object to consider is not the off-shell 2-point function, but rather the on-shell Compton amplitude
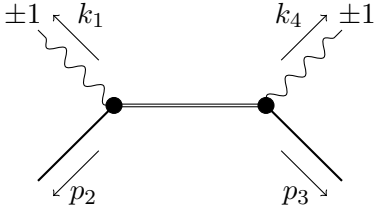
\begin{center}
    \begin{tikzpicture}
        \draw [-,thick] (-1,-1)--(0,0);
        \draw [-, double] (0,0)--(2,0);
        \filldraw (0,0) circle (3pt);
        \filldraw (2,0) circle (3pt);
        \draw [-,thick] (2,0)--(3,-1);
        \draw [vector] (0,0)--(-1,1);
        \draw [vector] (2,0)--(3,1);
        \draw[<-] (-0.8,1.2)--(-0.2,0.6);
        \draw[<-] (-0.8,-1.2)--(-0.2,-0.6);
        \draw[->] (2.2,0.6)--(2.8,1.2);
        \draw[->] (2.2,-0.6)--(2.8,-1.2);
        \node at (-0.3,1.2) {$k_1$};
        \node at (-0.4,-1.2) {$p_2$};
        \node at (2.3,1.2) {$k_4$};
        \node at (2.4,-1.2) {$p_3$};
        \node at (-1.2,1.2) {$\pm 1$};
        \node at (3.2,1.2) {$\pm 1$};
    \end{tikzpicture}
\end{center}
\begin{align}
    \mathcal{M}^{(\gamma)}_{+-} &=-\frac{4\kappa^2}{m_2^2}
[1|p_2 |4\rangle^2\int_{m_2^2}^\infty\mathrm{d}\mu^2\frac{\rho_{1}\left(\mu^2\right)}{(k_1+p_2)^2-\mu^2+\imath 0} \;\;\;+ \;\;\;\left(2\leftrightarrow 3\right), \nonumber\\
\mathcal{M}^{(\gamma)}_{++} &= \mathcal{M}^{(\gamma)}_{--} = 0.
\end{align}
From this on-shell perspective, we see that the somewhat abstruse electromagnetic self-duality constraint (\ref{emduality}) manifests as the vanishing of helicity violating Compton amplitudes \cite{Rosly:2002jt,Agullo:2016lkj,Novotny:2018iph,Agullo:2018nfv,Agullo:2018iya}. We then proceed as in the scalar case above, calculating the imaginary part of the forward Compton amplitude leading to the absorption cross section
\begin{equation}
    \sigma^{(\gamma)}_{\text{abs}}\left(\omega\right) = \frac{2\pi \kappa^2 \omega}{m_2} \rho_1\left(m_2^2+2m_2 \omega\right).
\end{equation}
Matching this with the UV cross section (\ref{sigmaUV}) gives the near-threshold spectral density
\begin{equation}
    \label{rho1}
    \rho_1\left(\mu^2\right) \sim \frac{G_\mathrm{N}^3 m_2^4 }{6\pi} \left(\mu^2-m_2^2\right), \hspace{5mm}\text{as} \hspace{5mm} \mu^2 \rightarrow m_2^2.
\end{equation}

\subsubsection*{Graviton Absorption Cross Section}

For graviton absorption, the analysis of the relevant 2-point function 
\begin{equation}
    \langle \mathcal{O}_2^{\mu_1\mu_2\mu_3\mu_4}(x) \mathcal{O}_2^{\nu_1\nu_2\nu_3\nu_4}(0)\rangle_X = \int \hat{\text{d}}^D k \; e^{\imath k\cdot x} \int_{m_2^2}^\infty \text{d}\mu^2 \frac{\imath \Pi_2^{\mu_1\mu_2\mu_3\mu_4\nu_1\nu_2\nu_3\nu_4}\left(\mu^2,k\right)}{k^2-\mu^2+\imath 0}.
\end{equation}
is somewhat more involved due to the proliferation of Lorentz indices. Without loss of generality we can assume that this operator has the same symmetry and tracelessness properties as the Weyl tensor 
\begin{equation}
    \label{O2symmetries}
    \mathcal{O}_2^{\mu\nu\rho\sigma} = - \mathcal{O}_2^{\nu\mu\rho\sigma}, \hspace{5mm} \mathcal{O}_2^{\mu\nu\rho\sigma} = \mathcal{O}_2^{\rho\sigma\mu\nu}, \hspace{5mm} \mathcal{O}_2^{\mu\nu\rho\sigma} + \mathcal{O}_2^{\mu\rho\sigma\nu} + \mathcal{O}_2^{\mu\sigma\nu\rho} = 0, \hspace{5mm} \eta_{\mu\rho}\mathcal{O}_2^{\mu\nu\rho\sigma} = 0. 
\end{equation}
In addition, as for photon absorption, the 2-point function is further constrained by a hidden \textit{gravitational self-duality} symmetry. The symmetry in this case is the duality invariance of the \textit{linearized} Einstein equations \cite{Henneaux:2004jw} on a Petrov type-D background \cite{Porto:2007qi}. The implications of this symmetry for black hole perturbation theory, in particular the quasi-normal mode isospectrality of axial and polar perturbations of Schwarzschild black holes was first pointed out by Chandrasekhar \cite{Chandrasekhar:1975nkd}. Following \cite{Goldberger:2005cd} we impose self-duality as a constraint on the 2-point function in the form
\begin{equation}
    \label{O2duality}
    \Pi_2^{\mu_1\mu_2\mu_3\mu_4\nu_1\nu_2\nu_3\nu_4}\left(\mu^2,k\right) = \frac{1}{4}{\epsilon^{\mu_1 \mu_2}}_{\alpha_1 \alpha_2}{\epsilon^{\nu_1 \nu_2}}_{\beta_1 \beta_2}\Pi_2^{\alpha_1\alpha_2\mu_3\mu_4\beta_1\beta_2\nu_3\nu_4}\left(\mu^2,k\right).
\end{equation}
We find that there is a unique solution to the combined constraints (\ref{O2symmetries}) and (\ref{O2duality}), the somewhat complicated expression is given in (\ref{gravitonnumerator1}) and (\ref{gravitonnumerator2}). As for the case of photon absorption above, the natural on-shell object to consider is the graviton Compton amplitude \\
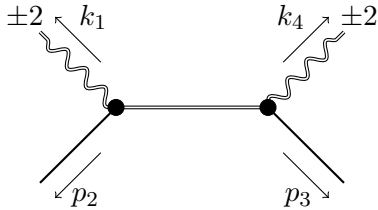
\begin{center}
    \begin{tikzpicture}
        \draw [-,thick] (-1,-1)--(0,0);
        \draw [-, double] (0,0)--(2,0);
        \draw [-,thick] (2,0)--(3,-1);
        \draw [vector,double] (0,0)--(-1,1);
        \draw [vector,double] (2,0)--(3,1);
        \draw[<-] (-0.8,1.2)--(-0.2,0.6);
        \draw[<-] (-0.8,-1.2)--(-0.2,-0.6);
        \draw[->] (2.2,0.6)--(2.8,1.2);
        \draw[->] (2.2,-0.6)--(2.8,-1.2);
        \node at (-0.3,1.2) {$k_1$};
        \node at (-0.4,-1.2) {$p_2$};
        \node at (2.3,1.2) {$k_4$};
        \node at (2.4,-1.2) {$p_3$};
        \node at (-1.2,1.2) {$\pm 2$};
        \node at (3.2,1.2) {$\pm 2$};
        \filldraw (0,0) circle (3pt);
        \filldraw (2,0) circle (3pt);
    \end{tikzpicture}
\end{center}
\begin{align}
    \mathcal{M}^{(h)}_{+-} &= -\frac{64\kappa^2}{m_2^4}[1|p_2 |4\rangle^4\int_{m_2^2}^\infty \mathrm{d}\mu^2\frac{\rho_{2}\left(\mu^2\right)}{(k_1+p_2)^2-\mu^2+\imath 0} \;\;\;+ \;\;\;\left(2\leftrightarrow 3\right), \nonumber\\
\mathcal{M}^{(h)}_{++} &= \mathcal{M}^{(h)}_{--} = 0,
\end{align}
where the vanishing of the helicity violating amplitudes is again the on-shell manifestation of self-duality. From this expression we calculate the absorption cross section 
\begin{equation}
    \sigma^{(h)}_{\text{abs}}\left(\omega\right) = 32\pi \kappa^2 m_2^2 \omega^3 \rho_2\left(m_2^2+2m_2 \omega\right),
\end{equation}
matching this with the UV cross section (\ref{sigmaUV}) gives the near-threshold spectral density
\begin{equation}
    \label{rho2}
    \rho_2\left(\mu^2\right) \sim \frac{G_\mathrm{N}^5 m_2^6 }{360\pi} \left(\mu^2-m_2^2\right), \hspace{5mm}\text{as} \hspace{5mm} \mu^2 \rightarrow m_2^2.
\end{equation}

\section{Absorptive Impulse}
\label{sec:impulse}

\subsection{In-In Scattering Observables and KMOC }
\label{subsec:KMOC}

Our goal is now, in the spirit of EFT, to recycle the information contained in the near-threshold expanded spectral functions (\ref{rho0}), (\ref{rho1}), (\ref{rho2}) to calculate distinct low-energy observables. Specifically, we calculate the leading contribution to the impulse on a Schwarzschild black hole scattering with a second compact body sourcing a scalar, electromagnetic or gravitational field. In this paper we approach this problem from an on-shell scattering amplitudes perspective, where the natural in-in formalism was formulated by KMOC \cite{Kosower:2018adc}. 

We begin with an executive summary of the KMOC formalism, see \cite{Kosower:2018adc,Herrmann:2021tct} for a more comprehensive description. In a general quantum mechanical system admitting an S-matrix describing time evolution from $t=-\infty$ to $t=+\infty$, the asymptotic change in the expectation value of an observable $\mathbb{O}$, initially in some state $|\text{in}\rangle$, is given by the formal expression
\begin{equation}    
    \label{KMOCv1}
    \Delta O = \langle \text{in}|S^\dagger \mathbb{O} S |\text{in}\rangle - \langle \text{in}|\mathbb{O}|\text{in}\rangle.
\end{equation}
In the present context, $S$ corresponds to the S-matrix on the \textit{complete} Hilbert space $\mathcal{H}_{\text{vis}}\otimes \mathcal{H}_{\text{inv}}$ and is therefore unitary: $S^\dagger S =1$. By making the standard definition $S\coloneqq1+\imath  T$ and using the unitarity relation
\begin{equation}
    \label{unitarity}
    T^\dagger = T-\imath T^\dagger T,
\end{equation}
the above can be rewritten as
\begin{equation}
    \label{KMOCv2}
    \Delta O = \imath \langle \text{in}|[\mathbb{O},T]|\text{in}\rangle + \langle \text{in}|T^\dagger [\mathbb{O},T]|\text{in}\rangle,
\end{equation}
where the two terms on the right-hand-side are sometimes referred to as the \textit{virtual} and \textit{real} contributions respectively. The different representations of the KMOC formula have complementary virtues: (\ref{KMOCv1}) manifests the fact that if $\mathbb{O}$ is Hermitian then the change in the expectation value is real-valued, while (\ref{KMOCv2}) manifests the fact that if $\mathbb{O}$ is a generator of a symmetry and commutes with $T$ then the expectation value is time independent.

We will now apply this formalism to calculate the asymptotic \textit{impulse} $\Delta p_2^\mu$ of a black hole undergoing 2-to-2 scattering with a second compact body. The in-state is chosen to be 
\begin{equation}
    |\text{in}\rangle \coloneqq\int \hat{\text{d}}^D p_1 \hat{\text{d}}^D p_2 \hat{\delta}^{(+)}\left(p_1^2-m_1^2\right) \hat{\delta}^{(+)}\left(p_2^2-m_2^2\right) \psi_{1}(p_1)\psi_{2}(p_2) e^{-\imath  p_1\cdot b_1}e^{- \imath p_2\cdot b_2} |p_1,p_2\rangle,
\end{equation}
where the wave-packets $\psi_{i}$, defined explicitly in \cite{Kosower:2018adc}, are chosen to localize particle-$i$ around the classical free particle trajectory with 4-velocity $u_i^\mu$ and impact parameter $b_i^\mu$. In the $\hbar\rightarrow 0$ limit, the quantum mechanical uncertainty in position and momentum vanishes and the corresponding asymptotic change in the expectation value of the operator $\mathbb{P}_2^\mu$ reduces to the corresponding \textit{classical} impulse. To extract the classical contributions efficiently, the resulting expressions are expanded before loop integration using the method of regions \cite{Beneke:1997zp}. In particular we expand to leading non-trivial order in the \textit{soft region} \cite{Herrmann:2021tct} defined by the scaling
\begin{equation}
    \label{soft}
    q^\mu \sim \ell^\mu \ll u_i^\mu \sim m_i.
\end{equation}
After simplifying \cite{Kosower:2018adc}, the KMOC formula for the classical impulse takes the form 
\begin{equation}
    \label{p2KMOC}
    \Delta p_2^\mu = \frac{1}{4m_1 m_2}\int \hat{\text{d}}^D q \;\hat{\delta}\left(u_1\cdot q\right) \hat{\delta}\left(u_2\cdot q\right) e^{\imath q\cdot b} \left[\mathcal{I}_v^\mu + \mathcal{I}_r^\mu\right],
\end{equation}
where $b^\mu \coloneqq b_1^\mu-b_2^\mu$ is the relative impact parameter, without loss of generality defined to satisfy $u_i\cdot b = 0$. The functions $\mathcal{I}^\mu_{v,r}$ are referred to as the virtual/real KMOC \textit{kernels}, corresponding to the respective terms in (\ref{KMOCv2}). The virtual kernel is straightforwardly related to an elastic scattering amplitude
\begin{equation}
    \label{virtualkernel}
    \mathcal{I}^\mu_{v} =  -\imath q^\mu \mathcal{M}\left(\phi_1(m_1 u_1) \phi_2(m_2 u_2) \rightarrow \phi_1(m_1 u_1+q) \phi_2(m_2 u_2-q)\right).
\end{equation}
The real kernel is given by inserting a complete set of states in the second term of (\ref{KMOCv2}). Restricting to the black hole ground state
\begin{equation}
    \mathds{1} \supset \int \hat{\mathrm{d}}^D r_1 \hat{\mathrm{d}}^D r_2 \;\hat{\delta}^{(+)}\left(r_1^2-m_1^2\right)\hat{\delta}^{(+)}\left(r_2^2-m_2^2\right) |\phi_1(r_1)\phi_2(r_2)\rangle \langle \phi_1(r_1)\phi_2(r_2)|,
\end{equation}
gives the \textit{conservative} contribution. In this paper we are interested in the leading absorptive contribution, this corresponds to an insertion of the $X$-states
\begin{equation}
    \mathds{1} \supset \sum_i\int_{m_2^2}^\infty \text{d}\mu^2\rho_i\left(\mu^2\right)\int \hat{\mathrm{d}}^D r_1 \hat{\mathrm{d}}^D r_2 \;\hat{\delta}^{(+)}\left(r_1^2-m_1^2\right)\hat{\delta}^{(+)}\left(r_2^2-\mu^2\right) |\phi_1(r_1)X(r_2)\rangle \langle \phi_1(r_1)X(r_2)|,
\end{equation}
where the sum over $i$ includes all internal quantum numbers of the excited states including spin. The resulting contribution to the real kernel is then given by 
\begin{align}   
    \mathcal{I}^\mu_r &= \sum_i\int_{m_2^2}^\infty \text{d}\mu^2\rho_i\left(\mu^2\right) \int \hat{\text{d}}^D \ell \;\hat{\delta}^{(+)}\left((m_1 u_1-\ell)^2-m_1^2\right) \hat{\delta}^{(+)}\left((m_2 u_2+\ell)^2-\mu^2\right)\\
    &\hspace{42mm}\times \ell^\mu \times \mathcal{M}\left(\phi_1(m_1 u_1)\phi_2(m_2 u_2)\rightarrow \phi_1(m_1 u_1-\ell)X(m_2 u_2+\ell)\right) \nonumber\\
    &\hspace{50mm}\times\mathcal{M}^*\left(\phi_1(m_1 u_1+q)\phi_2(m_2 u_2-q)\rightarrow \phi_1(m_1 u_1-\ell)X(m_2 u_2+\ell)\right).\nonumber
\end{align}
Since we are interested in the leading-order PM contribution to the absorptive impulse we can make use of the unitarity relation (\ref{unitarity}) to replace the conjugated amplitude $\mathcal{M}^*$ with the unconjugated time-reversed process. This expression can be rewritten in the compact diagrammatic weighted cut notation defined in Appendix \ref{app:cuts}:
\begin{center}
    \begin{tikzpicture}
        \node at (-5,0) {};
        \node at (-2,-1) {$\mathcal{I}^\mu_{r}$};
        \node at (-1,-1) {$=$};
        \draw[-,thick] (0,0)--(2,0);
        \draw[-,double] (2,0)--(3,0);
        \draw[vector,double] (2,0)--(2,-2);
        \draw[fermion] (0,-2)--(2,-2);
        \draw[fermion] (2,-2)--(3,-2);
        \draw[dashed,thick,red] (3,0.5)--(3,-0.6);
        \draw[dashed,thick,red] (3,-1.4)--(3,-2.5);
        \draw[-,double] (3,0)--(4,0);
        \draw[-,thick] (4,0)--(6,0);
        \draw[vector,double] (4,0)--(4,-2);
        \draw[fermion] (3,-2)--(4,-2);
        \draw[fermion] (4,-2)--(6,-2);
        \draw[->] (0.5,0.25)--(1.5,0.25);
        \draw[->] (4.5,0.25)--(5.5,0.25);
        \draw[->] (1.65,-1.5)--(1.65,-0.5);
        \node at (1,0.5) {$m_2 u_2$};
        \node at (1,-2.3) {$m_1 u_1$};
        \node at (5,0.5) {$m_2 u_2-q$};
        \node at (1.45,-1) {$\ell$};
        \filldraw (2,0) circle (3pt);
        \filldraw (4,0) circle (3pt);
        \node at (3.1,-1) {\color{red}$-\ell^\mu$\color{black}};
        \node at (6.5,-1) {.};
        \node at (8.5,-1) {
        \begin{minipage}{3cm}
          \begin{equation}
          \label{realkernel}
              \;
          \end{equation}  
        \end{minipage}
        };
    \end{tikzpicture}
\end{center}
Finally, at leading order there is no mixing between radiation and horizon absorption; the leading absorptive impulse on the small body can then be obtained trivially using conservation of momentum
\begin{equation}
    \label{deltap1deltap2}
    \left(\Delta p_1^\mu\right)_{\text{abs}} = - \left(\Delta p_2^\mu\right)_{\text{abs}}.
\end{equation}

\subsection{Heavy Modes and Love Numbers }
\label{subsec:Love}

By assumption the state created by $\phi_2$ is a stable ground state, meaning $\rho\left(\mu^2<m_2^2\right) = 0$. Before proceeding it is therefore useful to define a shifted spectral parameter and spectral density
\begin{equation}
\label{defs}
    \rho\left(\mu^2\right) \coloneqq\tilde{\rho}(s), \hspace{10mm} s \coloneqq\frac{\mu^2 -m_2^2}{2m_2},
\end{equation}
where $\tilde{\rho}(s<0) =0$. In this notation, both the virtual (\ref{virtualkernel}) and real (\ref{realkernel}) absorptive contributions to the KMOC formula (\ref{p2KMOC}) require evaluating a spectral integral from $0 < s< \infty$. Naively this is problematic as it would require detailed knowledge of the spectral function $\tilde{\rho}\left(s\right)$ at all energy scales. However, there is clearly an important distinction between those $X$-states that can go on-shell during the scattering event and those that cannot. To illustrate the distinction in a physically transparent manner, we introduce a cutoff $\Lambda$, splitting the spectral integral into two contributions:
\begin{align}
    &\textit{Light modes:} \hspace{10mm} 0<s<\Lambda \hspace{5mm} \Leftrightarrow \hspace{5mm}  s\sim q\nonumber\\
    &\textit{Heavy modes:} \hspace{8mm} \Lambda<s<\infty \hspace{3.5mm} \Leftrightarrow \hspace{5mm} s\gg q,
\end{align}
where $q$ is the momentum transfer, assumed to be small in the soft expansion (\ref{soft}). In the calculation of the virtual kernel (\ref{virtualkernel}), the contribution to the $X$-state propagator from the heavy modes can never be on-shell; expanding to leading order in the soft region
\begin{equation}
    \label{pinch}
    \int_\Lambda^\infty\text{d}s\;\frac{\tilde{\rho}\left(s\right)}{2m_2(u_2\cdot \ell) +\ell^2-2m_2 s+\imath 0} \approx -\frac{1}{2m_2}\int_\Lambda^\infty\text{d}s \;\frac{\tilde{\rho}\left(s\right)}{s}.
\end{equation}
\newcommand{\ctOneLoopV}{
\begin{tikzpicture}[scale=1]
		\coordinate (e1) at (-2,-1);
		\coordinate (e2) at (-2,1);
		\coordinate (e3) at (2,1);
		\coordinate (e4) at (2,-1);
		
		\coordinate (v1) at (0,1);
        \coordinate (v2) at (-1,-1);
        \coordinate (v3) at (1,-1);
		
		\draw[thick] (e2) -- (v1) -- (e3) ;
        \draw[fermion] (e1) -- (e4) ;
		\draw[dashed](v2) -- (v1) -- (v3) ;

        \path[draw=black,line width=1.5pt, fill=white] (v1) circle[radius=0.25];
        \path[draw=black, line width=1.5pt] (v1) -- +(0.17,0.17);
        \path[draw=black, line width=1.5pt] (v1) -- +(-0.17,-0.17);
        \path[draw=black, line width=1.5pt] (v1) -- +(0.17,-0.17);
        \path[draw=black, line width=1.5pt] (v1) -- +(-0.17,0.17);
\end{tikzpicture}}
That is, we find that for the heavy modes the $X$-state propagator \textit{pinches}, generating an effective contact diagram shown in figure~\ref{fig:ContactDiagram}.
\begin{figure}
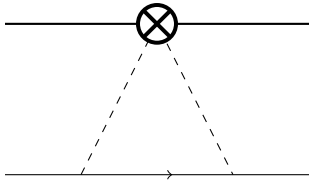

    \centering
    \ctOneLoopV
    \caption{Effective contact vertex obtained from the heavy modes in the KL representation.}
    \label{fig:ContactDiagram}
\end{figure}
The contact term can be interpreted as a new effective operator contribution to the visible sector action (\ref{Svis}). For example, for massless scalar absorption the leading-order operator takes the form
\begin{equation}
    S_{\text{vis}} \supset \int \text{d}^4 x \sqrt{-g}\left[c \psi^2 \phi_2^2 + ...\right], \hspace{5mm} \text{where} \hspace{5mm} c \sim \int_\Lambda^\infty\text{d}s \;\frac{\tilde{\rho}\left(s\right)}{s}.
\end{equation}
The Wilson coefficients of such higher-dimension operators are, in this context, usually referred to as \textit{tidal Love numbers} \cite{Binnington:2009bb,Cheung:2020sdj,Bern:2020uwk}. Clearly, we should never have included the heavy modes to begin with, they are always off-shell and so can be consistently integrated out of the point-particle effective description. We will therefore assume that this has been done and the associated Love number contributions to observables calculated as part of the conservative contribution \cite{Cheung:2020sdj,Bern:2020uwk}. 

By contrast, the light modes (gapless modes in the language of Section \ref{sec:intro}) cannot be integrated out since they can go on-shell during the scattering event. To calculate their contribution to the impulse we only need the spectral function in a small neighbourhood of the threshold value $s\rightarrow 0$, and as demonstrated in Section \ref{subsec:crosssection}, this information can obtained by matching the absorption cross section expanded around $\omega \rightarrow 0$. 

In practice the crude (though physically transparent) cutoff $\Lambda$ can be replaced by an analytic regularization of the spectral integral for the light modes
\begin{equation}
    \int_0^\Lambda \text{d}s \;\tilde{\rho}(s) \hspace{5mm} \Rightarrow \hspace{5mm} \int_0^\infty \text{d}s \; s^\alpha\tilde{\rho}(s).
\end{equation}
In this perspective the spectral integral should be taken together with the dimensionally regularized loop integral as the object to be expanded using the method of regions. The soft region is then defined by 
\begin{equation}
    \label{ssoft}
    s\sim q^\mu \sim \ell^\mu  \ll u_i^\mu \sim m_i,
\end{equation}
and therefore only the leading-order term in the Taylor expansion of $\tilde{\rho}(s)$ around $s=0$ contributes to leading-order in the soft expansion. By analogy with dimensional regularization, after expansion in the soft region the analytically regularized $s$-integration domain is extended to cover the entire range $0<s<\infty$, evaluated for convergent values of $\alpha$ and then analytically continued to $\alpha=0$. This form of regularization has many familiar advantages, scaleless $s$-integrals evaluate to zero, power-law UV divergences are absent and logarithmic UV divergences show up as finite order poles at $\alpha=0$. 

\subsection{Light Modes and Absorption }
\label{subsec:light}

To simplify the calculation of the absorptive impulse, it is useful to first decompose the kernel into scalar contributions
\begin{equation}
    \mathcal{I}^\mu = q^\mu\mathcal{I}_{q} + \check{u}_1^\mu\mathcal{I}_{\check{u}_1} + \check{u}_2^\mu\mathcal{I}_{\check{u}_2},
\end{equation}
where our kinematic conventions and notation are defined in Appendix \ref{app:conventions}. To calculate these contributions to the real kernel we decompose the loop momentum insertion in (\ref{realkernel}) as\footnote{The component perpendicular to the scattering plane integrates to zero.}
\begin{equation}
    \ell^\mu \rightarrow \left(\frac{\ell\cdot q}{q^2}\right) q^\mu + \left(u_1\cdot \ell\right)\check{u}_1^\mu + \left(u_2\cdot \ell\right)\check{u}_2^\mu.
\end{equation}
Because of the cut condition for particle-1, $u_1\cdot \ell=0$, we trivially find that 
\begin{equation}
    \mathcal{I}_{r,\check{u}_1} = 0.
\end{equation}
The remaining non-vanishing contributions $\mathcal{I}_{r,q}$ and $\mathcal{I}_{r,\check{u}_2}$ will be denoted \textit{transverse} and \textit{longitudinal} respectively. The transverse contribution can be simplified using
\begin{equation}
    \ell\cdot q = \frac{1}{2}\left(\ell+q\right)^2 -\frac{1}{2}\ell^2 - \frac{1}{2}q^2,
\end{equation}
the first two terms pinch mediator propagators and therefore produce vanishing scaleless integrals. For the longitudinal contribution we use the cut condition for particle-2, $u_2\cdot \ell = s$, where we use the shifted spectral parameter $s$ defined in (\ref{defs}). All together we find the effective decomposition 
\begin{equation}
    \ell^\mu \rightarrow -\frac{1}{2}q^\mu + s\check{u}_2^\mu,
\end{equation}
or diagrammatically:
\begin{center}
    \begin{tikzpicture}
        \node at (-5,0) {};
        \node at (-2,-1) {$\mathcal{I}_{r,q}$};
        \node at (-1,-1) {$=$};
        \draw[-,thick] (0,0)--(2,0);
        \draw[-,double] (2,0)--(3,0);
        \draw[vector,double] (2,0)--(2,-2);
        \draw[fermion] (0,-2)--(2,-2);
        \draw[fermion] (2,-2)--(3,-2);
        \draw[dashed,thick,red] (3,0.5)--(3,-0.6);
        \draw[dashed,thick,red] (3,-1.4)--(3,-2.5);
        \draw[-,double] (3,0)--(4,0);
        \draw[-,thick] (4,0)--(6,0);
        \draw[vector,double] (4,0)--(4,-2);
        \draw[fermion] (3,-2)--(4,-2);
        \draw[fermion] (4,-2)--(6,-2);
        \draw[->] (0.5,0.25)--(1.5,0.25);
        \draw[->] (4.5,0.25)--(5.5,0.25);
        \draw[->] (1.65,-1.5)--(1.65,-0.5);
        \node at (1,0.5) {$m_2 u_2$};
        \node at (1,-2.3) {$m_1 u_1$};
        \node at (5,0.5) {$m_2 u_2-q$};
        \node at (1.45,-1) {$\ell$};
        \filldraw (2,0) circle (3pt);
        \filldraw (4,0) circle (3pt);
        \node at (3,-1) {\color{red}$\frac{1}{2}$\color{black}};
        \node at (6.5,-1) {,};
        \node at (8.5,-1) {
        \begin{minipage}{3cm}
          \begin{equation}
              \;
          \end{equation}  
        \end{minipage}
        };
    \end{tikzpicture}
\end{center}
\begin{center}
    \begin{tikzpicture}
        \node at (-5,0) {};
        \node at (-2,-1) {$\mathcal{I}_{r,\check{u}_2}$};
        \node at (-1,-1) {$=$};
        \draw[-,thick] (0,0)--(2,0);
        \draw[-,double] (2,0)--(3,0);
        \draw[vector,double] (2,0)--(2,-2);
        \draw[fermion] (0,-2)--(2,-2);
        \draw[fermion] (2,-2)--(3,-2);
        \draw[dashed,thick,red] (3,0.5)--(3,-0.6);
        \draw[dashed,thick,red] (3,-1.4)--(3,-2.5);
        \draw[-,double] (3,0)--(4,0);
        \draw[-,thick] (4,0)--(6,0);
        \draw[vector,double] (4,0)--(4,-2);
        \draw[fermion] (3,-2)--(4,-2);
        \draw[fermion] (4,-2)--(6,-2);
        \draw[->] (0.5,0.25)--(1.5,0.25);
        \draw[->] (4.5,0.25)--(5.5,0.25);
        \draw[->] (1.65,-1.5)--(1.65,-0.5);
        \node at (1,0.5) {$m_2 u_2$};
        \node at (1,-2.3) {$m_1 u_1$};
        \node at (5,0.5) {$m_2 u_2-q$};
        \node at (1.45,-1) {$\ell$};
        \filldraw (2,0) circle (3pt);
        \filldraw (4,0) circle (3pt);
        \node at (3,-1) {\color{red}$-s$\color{black}};
        \node at (6.5,-1) {.};
        \node at (8.5,-1) {
        \begin{minipage}{3cm}
          \begin{equation}
              \;
          \end{equation}  
        \end{minipage}
        };
    \end{tikzpicture}
\end{center}
We can re-express these in a compact form by defining a \textit{partial} amplitude $\mathcal{M}_4\left(s,q^2\right)$ corresponding to the contributions of $X$-states with fixed invariant mass to the elastic amplitude $\mathcal{M}\left(q^2\right)$; by definition
\begin{equation}
    \mathcal{M}_4\left(q^2\right) = \int_{m_2^2}^\infty \text{d}\mu^2 \; \rho\left(\mu^2\right) \mathcal{M}_4\left(\mu^2,q^2\right) = 2m_2\int_0^\infty \text{d}s \; \tilde{\rho}(s) \mathcal{M}_4\left(s,q^2\right).
\end{equation}
The transverse and longitudinal contributions to the real kernel then take the form
\begin{align}
    \label{kernelrealscalar}
     \mathcal{I}_{r,q} = -\text{Im}\mathcal{M}_4\left(q^2\right), \hspace{10mm}
     \mathcal{I}_{r,\check{u}_2} = 4m_2\int_0^\infty \text{d}s\;s\;\tilde{\rho}(s)\;\text{Im}\mathcal{M}_4\left(s,q^2\right).
\end{align}
The amplitudes appearing in these expressions are assumed to be expanded to leading order in the generalized soft limit (\ref{ssoft}). In the corresponding one-loop contribution to the conservative impulse, the leading soft contribution is \textit{super-classical} and cancels when the real and virtual kernels are summed \cite{Kosower:2018adc}. In the absorptive case however, the leading soft contribution is the classical contribution due to an $\hbar/\hbar$ cancellation; the super-classical scaling of the box diagram is compensated by the ``quantum" scaling of the $X$-state effective propagator.  

Next we consider the virtual kernel (\ref{virtualkernel}), which has only a transverse contribution. Decomposing this into real and imaginary parts
\begin{equation}
    \label{kernelvirtualscalar}
    \mathcal{I}_{v,q} = \text{Im}\mathcal{M}_4\left(q^2\right)- \imath\, \text{Re}\mathcal{M}_4\left(q^2\right),
\end{equation}
we find that the transverse part of the real kernel is exactly cancelled by a corresponding contribution from the virtual kernel\footnote{Each of these pieces individually produce \textit{imaginary} contributions to the impulse after Fourier transforming; this cancellation is therefore a consequence of the non-manifest reality of the observable.}. 

\begin{figure}
    \centering
\begin{tikzpicture}
        \begin{scope}
        \draw[-,thick] (0,0)--(2,0);
        \draw[-,double] (2,0)--(3,0);
        \draw[vector,double] (2,0)--(2,-2);
        \draw[fermion] (0,-2)--(2,-2);
        \draw[fermion] (2,-2)--(4,-2);
        \draw[-,double] (3,0)--(4,0);
        \draw[-,thick] (4,0)--(6,0);
        \draw[vector,double] (4,0)--(4,-2);
        \draw[fermion] (4,-2)--(6,-2);
        \draw[->] (0.5,0.25)--(1.5,0.25);
        \draw[->] (4.5,0.25)--(5.5,0.25);
        \draw[->] (1.65,-1.5)--(1.65,-0.5);
        \node at (1,0.5) {$m_2 u_2$};
        \node at (1,-2.3) {$m_1 u_1$};
        \node at (5,0.5) {$m_2 u_2-q$};
        \node at (1.45,-1) {$l$};
        \filldraw (2,0) circle (3pt);
        \filldraw (4,0) circle (3pt);
        \end{scope}
        \begin{scope}[xshift=7cm]
        \draw[-,thick] (0,0)--(2,0);
        \draw[-,double] (2,0)--(3,0);
        \draw[vector,double] (2,0)--(4,-2);
        \draw[fermion] (0,-2)--(2,-2);
        \draw[fermion] (2,-2)--(4,-2);
        \draw[-,double] (3,0)--(4,0);
        \draw[-,thick] (4,0)--(6,0);
        \draw[vector,double] (4,0)--(2,-2);
        \draw[fermion] (4,-2)--(6,-2);
        \draw[->] (0.5,0.25)--(1.5,0.25);
        \draw[->] (4.5,0.25)--(5.5,0.25);
        \draw[->] (2.65,-1)--(2,-0.4);
        \node at (1,0.5) {$m_2 u_2$};
        \node at (1,-2.3) {$m_1 u_1$};
        \node at (5,0.5) {$m_2 u_2-q$};
        \node at (2.1,-1) {$l$};
        \filldraw (2,0) circle (3pt);
        \filldraw (4,0) circle (3pt);
        \end{scope}
    \end{tikzpicture}
    \caption{The box and the crossed-box diagram are the two Feynman diagrams necessary for the computation of the leading order mass-shift. Wiggly lines represent massless force carriers: scalars, photons or gravitons.}
    \label{fig:boxDiagrams}
\end{figure}
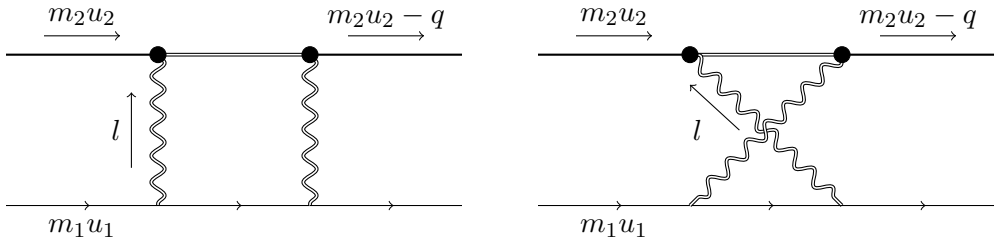

Combining (\ref{kernelrealscalar}) and (\ref{kernelvirtualscalar}) we find that the transverse impulse (proportional to $b^\mu$) receives contributions only from the real part of the elastic amplitude. In general this takes the form 
\begin{equation}
    \label{elasticgeneralform}
    \mathcal{M}_4\left(q^2\right) = \int \hat{\text{d}}^D \ell \; \mathcal{N}\left[(u_2\cdot \ell),q^2,y\right] \frac{\hat{\delta}\left(u_1\cdot \ell\right)}{\ell^2(\ell+q)^2}\int_0^\infty \text{d}s\; \frac{s^{1+\alpha}}{u_2\cdot \ell -s + \imath 0},
\end{equation}
where $\mathcal{N}$ is a model dependent polynomial and as above we have dropped scaleless contributions that pinch mediator propagators. The cut propagator for particle-1 arises from interference between the box and crossed-box diagrams in the soft limit, which are shown in Figure~\ref{fig:boxDiagrams}.
using the distributional identity
\begin{equation}
    \frac{1}{u_1\cdot \ell - \imath 0} - \frac{1}{u_1\cdot \ell + \imath 0} = \imath\hat{\delta}\left(u_1\cdot \ell\right).
\end{equation}
Importantly, since we have expanded to leading-order in the soft limit (\ref{ssoft}) the $\mathcal{N}$ polynomial is homogeneous
\begin{equation}
    \mathcal{N}\left[\lambda(u_2\cdot \ell),\lambda^2 q^2,y\right] = \lambda^k \mathcal{N}\left[(u_2\cdot \ell),q^2,y\right],
\end{equation}
where $k=0$ for scalars, $k=2$ for photons and $k=4$ for gravitons. Since $k$ is even in each case, $\mathcal{N}$ must only contain even powers of $(u_2\cdot \ell)$ and is therefore \textit{symmetric} under $\ell \rightarrow -\ell-q$. This becomes significant once we restrict to the real part of the spectral integral given by the Cauchy principal value (PV); in the analytic regularization described in Section \ref{subsec:Love} we calculate\footnote{We can alternatively calculate this integral using the cutoff regularization described in Section \ref{subsec:Love}. This leads to an additional \textit{power law} divergent contribution that is even under $\ell \rightarrow -\ell-q$ and therefore non-vanishing after loop integration. Physically, we interpret this as an ambiguity in the definition of the scale $\Lambda$ separating the light and heavy modes in the spectral integral. Importantly, since the \textit{logarithmically} divergent contributions vanish after loop integration, there is no associated classical RG running of the leading (static) Love numbers and therefore no tension with their observed vanishing in $d=4$ and recently discovered associated symmetries \cite{Charalambous:2022rre,Ivanov:2022qqt,Charalambous:2023jgq,Saketh:2023bul,Hui:2020xxx,Hui:2021vcv,Hui:2022vbh}.  } 
\begin{equation}
    \text{PV}\int_0^\infty \text{d}s\; \frac{s^{1+\alpha}}{u_2\cdot \ell -s }  = \frac{u_2\cdot \ell}{\alpha} + (u_2\cdot \ell) \log\left|u_2\cdot \ell\right|+\mathcal{O}\left(\alpha\right).
\end{equation}
We therefore find, for the contribution of the light modes, the real part of the integrand (\ref{elasticgeneralform}) is odd under the change of variables $\ell\rightarrow -\ell-q$ and integrates to zero. Note that, as discussed in Section \ref{subsec:Love}, the contributions of the heavy modes pinch the propagator (\ref{pinch}) giving an expression that is even under this change of variables and therefore a non-vanishing real part.

All together we conclude that, to leading non-trivial PM order, the absorptive contribution to the impulse is purely longitudinal, while the conservative contribution (including Love number operators) is purely transverse. The leading absorptive impulse is therefore completely determined by the \textit{mass-shift}, the change in the black hole rest mass during scattering, explicitly
\begin{equation}
    \label{deltaptodeltam}
    \left(\Delta p_2^\mu\right)_{\text{abs}} = \left(\Delta m_2 \right) \check{u}_2^\mu,
\end{equation}
where 
\begin{equation}
    \label{massshiftformula}
    \Delta m_2  = \frac{1}{m_1}\int \hat{\text{d}}^D q \;\hat{\delta}\left(u_1\cdot q\right)\hat{\delta}\left(u_2\cdot q\right)e^{\imath q\cdot b} \int_0^\infty\text{d}s\;s \; \tilde{\rho}(s) \; \text{Im}\mathcal{M}_4\left(s,q^2\right).
\end{equation}
In \cite{Goldberger:2020wbx} it was noted that, since for Schwarzchild black holes the area of the event horizon is $A=16\pi G_\mathrm{N}^2 m_2^2$, a decrease in mass during scattering would imply a violation of the Hawking area theorem \cite{Hawking:1971vc}. An appealing feature of the above simple formula is that it manifests the positivity of the mass-shift. Unitarity relates the imaginary part of the partial amplitude to the strictly positive absorptive cross section where the final $X$-state has fixed invariant mass. Together with the fact that the spectral density is non-negative and, crucially, vanishes for $\mu^2 < m_2^2$, positivity of the mass-shift is a trivial corollary. 

\subsection{Calculation of the Mass-Shift}
\label{subsec:mass}

In the explicit cases of scalar, photon and graviton absorption, the mass-shift can now be straightforwardly calculated from the imaginary parts of the relevant box diagrams depicted in Figure \ref{fig:boxDiagrams}. These can be obtained without difficulty either using the Feynman rules in Appendix \ref{app:Feynman} or, as discussed in more detail in Section \ref{subsec:neutron}, by directly sewing together the triangle cut using the Compton amplitudes calculated in Section \ref{subsec:crosssection}. There are then three remaining integrals to calculate ($\int\text{d}s$, $\int\text{d}^D \ell$ and $\int\text{d}^D q$), the necessary master integrals are collected in Appendix \ref{app:integrals}. 
The final results for the mass-shifts in each case are summarized in the following paragraphs; details on the notation conventions are given in Appendix \ref{app:conventions}.

\subsubsection*{Scalar Absorption }
\label{subsubsec:scalar}
The imaginary part of the box diagram for scalar absorption 
\begin{equation}
    \text{Im}\mathcal{M}_4^{(\psi)}\left(s,q^2\right) = \frac{4\pi G_\mathrm{N} Q_s^2}{ m_1 m_2} \int \hat{\text{d}}^D \ell\frac{\hat{\delta}\left(u_1\cdot \ell\right)\hat{\delta}\left(u_2\cdot \ell-s\right)}{\ell^2(\ell+q)^2},
\end{equation}
together with the near-threshold spectral density (\ref{rho0}) gives the corresponding mass-shift 
\begin{equation}
    \label{massshiftscalar}
    \boxed{\left(\Delta m_2\right)^{(\psi)} = \frac{G_\mathrm{N}^2 Q_s^2 m_2^2 }{32m_1^2 |b|^3}\sqrt{y^2-1}.}
\end{equation}

\subsubsection*{Photon Absorption }
\label{subsubsec:photon}
The imaginary part of the box diagram for photon absorption 
\begin{equation}
    \text{Im}\mathcal{M}_4^{(\gamma)}\left(s,q^2\right) = -\frac{16\pi G_\mathrm{N} Q_e^2 m_1 }{m_2} \int \hat{\text{d}}^D \ell\left[4s^2 +\left(2y^2-1\right) |q|^2\right] \frac{\hat{\delta}\left(u_1\cdot \ell\right)\hat{\delta}\left(u_2\cdot \ell-s\right)}{\ell^2(\ell+q)^2},
\end{equation}
together with the near-threshold spectral density (\ref{rho1}) gives the corresponding mass-shift
\begin{equation}
    \label{massshiftphoton}
    \boxed{\left(\Delta m_2\right)^{(\gamma)} = \frac{3G_\mathrm{N}^4 Q_e^2 m_2^4}{32 |b|^5}\left(5y^2-1\right)\sqrt{y^2-1}.}
\end{equation}

\subsubsection*{Graviton Absorption }
\label{subsubsec:graviton}
The imaginary part of the box diagram for graviton absorption 
\begin{align}
    \text{Im}\mathcal{M}_4^{(h)}\left(s,q^2\right) &= \frac{256\pi^2 G_\mathrm{N}^2 m_1^3}{m_2} \int \hat{\text{d}}^D \ell\left[16s^4+8\left(4y^2-1\right)s^2 |q|^2 +\left(8y^4-8y^2+1\right) |q|^4\right] \nonumber\\
    &\hspace{40mm}\times \frac{\hat{\delta}\left(u_1\cdot \ell\right)\hat{\delta}\left(u_2\cdot \ell-s\right)}{\ell^2(\ell+q)^2} ,
\end{align}
together with the near-threshold spectral density (\ref{rho2}) gives the corresponding mass-shift 
\begin{equation}
    \label{massshiftgraviton}
    \boxed{\left(\Delta m_2\right)^{(h)} = \frac{5\pi G_\mathrm{N}^7 m_1^2 m_2^6 }{16|b|^7}\left(21y^4 - 14y^2+1\right)\sqrt{y^2-1},}
\end{equation}
in perfect agreement with previous worldline EFT calculations \cite{Goldberger:2020wbx}.

\subsection{Neutron Stars and Dissipation Numbers}
\label{subsec:neutron}

The results of the previous section for Schwarzschild black hole absorption were obtained after explicitly matching with a UV cross section and exploiting the additional simplification of duality invariance as explained in Section \ref{subsec:crosssection}. For generic compact bodies, such as neutron stars, this UV information may not be available. In such a case this formalism is still predictive, since the leading-order absorption depends only on the near-threshold Taylor series coefficients of the spectral density, we can proceed in the spirit of \textit{bottom up} EFT and use these unknown coefficients as a parametrization of our ignorance. As usual, at higher-orders finding a non-redundant parametrization is highly non-trivial; on-shell amplitudes based approaches are very well-suited to this problem, directly providing gauge invariant and non-redundant expressions by construction (for a relevant example in the context of Standard Model EFT see e.g. \cite{Li:2020gnx}). 

\begin{figure}
\newcommand{\cutGeneral}{\begin{tikzpicture}
        \begin{scope}
                \filldraw[fill=none,draw=none] (1,1) rectangle (5,-3);
        \draw[-,thick] (1,0)--(3,0);
        \draw[vector,double] (3,0)--(2,-2) ;
        \draw[fermion] (1,-2)--(2,-2);
        \draw[fermion] (2,-2)--(4,-2);
        \draw[-,thick] (3,0)--(5,0);
        \draw[vector,double] (3,0)--(4,-2)  ;
        \draw[fermion] (4,-2)--(5,-2);
        \filldraw[fill=gray!10] (3,0) circle (15pt) node {$\mathcal{M}_{c}$};
        \draw[->] (0.5,0.25)--(1.5,0.25);
        \draw[->] (4.5,0.25)--(5.5,0.25);
        \node at (1,0.5) {$m_2 u_2$};
        \node at (1,-2.3) {$m_1 u_1$};
        \node at (5,0.5) {$m_2 u_2-q$};
        \draw[dashed,thick,red] (2.8,-1.2)--(2,-0.9);
        \draw[dashed,thick,red] (4,-0.9)--(3.2,-1.2);
        \draw[dashed,thick,red] (3,-1.6)--(3,-2.4);
        \draw[->] (2.5,-0.5)--(2.3,-0.9);
        \draw[->] (3.5,-0.5)--(3.7,-0.9);
        \node at (2.2,-0.5) {$k_1$};
        \node at (3.8,-0.5) {$k_4$};
        \end{scope}
    \end{tikzpicture}}
    \centering
    \cutGeneral
    \caption{The Compton cut is the only unitary cut needed in order to compute leading-order classical observables. In the explicit Compton amplitudes given below the mediator momenta are labelled as shown, after sewing we identify $k_1=-l$, $k_4=l+q$.  }
    \label{fig:ComptonCut}
\end{figure}
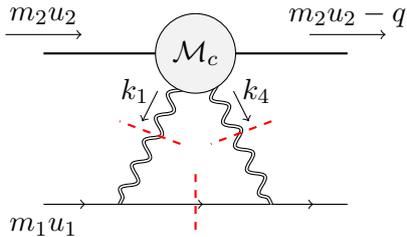

As we have seen explicitly, the input required to compute the mass-shift is the absorptive contribution to the 2-to-2 elastic amplitude. Since we are only interested in long-distance effects, this quantity can be determined using standard on-shell unitarity-based methods, by computing the triangle cut of the amplitude depicted in Figure~\ref{fig:ComptonCut}. Absorptive effects are parametrized by the most general allowed form of the Compton amplitude. This task is somewhat complicated by the fact that, due to the unfamiliar $s$-integrals, the absorptive amplitudes are non-analytic functions of the external kinematics. 

Based on the general discussion in Section \ref{subsec:crosssection}, it is convenient to define a \textit{Compton kernel} $\mathcal{M}_c(s)$ as the integrand of the (low-energy) spectral integral, where this quantity is expanded in the soft limit (\ref{ssoft}). Let us first consider the case of the scalar-force model. The Compton kernel for scattering a scalar off a black hole takes the general form 
\begin{align}
    \mathcal{M}^{(\psi)}(s) &= -\frac{1}{u_2\cdot k_1-s+\imath 0}\left[\kappa^2 s\,  c^{(\psi)} +\kappa^4s^2 c_2^{(\psi)}+\kappa^4 (k_1\cdot k_4 )s c_3^{(\psi)}+\dots\right]+ \left(1\leftrightarrow 4\right)\,,
\end{align}
where the ellipsis denotes terms with higher orders in $G_\mathrm{N}$. Notice that terms including $u_2\cdot k_{1}$ and $u_2\cdot k_{4}$ are absent, since  $u_2\cdot (k_1+k_4) = -\frac{k_1\cdot k_4}{m_2}$ and pinch terms correspond to vanishing scaleless $s$-integrals. The coefficients $c_i^{(\psi)}$ provide an on-shell, gauge invariant, definition of so-called \textit{dissipation numbers} \cite{Saketh:2023bul}. In the following we work to leading order and keep only the term proportional to $c^{(\psi)}$, higher-order terms may be important for sub-leading absorptive effects.

For gravity and electrodynamics there are two independent Compton amplitudes corresponding to helicity conserving and helicity violating interactions; accordingly we have to parametrize two kernels $\mathcal{M}_{+\mp}(s)$ independently. For electrodynamics the most general form is 
\begin{align}
    \mathcal{M}_{+-}^{(\gamma)}(s) &= - \frac{[1|u_2|4\rangle^2}{u_2\cdot k_1-s+\imath 0} \left[\kappa^2 s\,c_{+-}^{(\gamma)}+\dots\right]+ \left(1\leftrightarrow 4\right)\,,\label{eq:PhotonComptonPM}\\
    \mathcal{M}_{++}^{(\gamma)}(s) &= - \frac{ [14]^2}{u_2\cdot k_1-s+\imath 0}\left[\kappa^2 s\, c_{++}^{(\gamma)}+\dots\right]+ \left(1\leftrightarrow 4\right)\label{eq:PhotonComptonPP}\,.
\end{align}
The spinor prefactors are fixed by the little-group covariance of the amplitudes. Again, the ellipsis denotes higher orders in $G_\mathrm{N}$ and we will focus on the leading-order term. The analysis for the gravitational case is similar
\begin{align}
    \mathcal{M}_{+-}^{(h)}(s) &= - \frac{[1|u_2|4\rangle^4}{u_2\cdot k_1-s+\imath 0} \left[\kappa^2s\, c_{+-}^{(h)}+\dots\right]+ \left(1\leftrightarrow 4\right)\,,\label{eq:GravitonComptonPM}\\
    \mathcal{M}_{++}^{(h)}(s) &= - \frac{ [14]^4}{u_2\cdot k_1-s+\imath 0}\left[\kappa^2s\, c_{++}^{(h)}+\dots\right]+ \left(1\leftrightarrow 4\right)\label{eq:GravitonComptonPP}\,.
\end{align}
We see the clear simplification compared with the off-shell construction of $\mathcal{O}$ operators. For completeness, these on-shell expression can be mapped onto a specific set of operators that are explicitly given in Appendix~\ref{sec:ExplicitLagr}.

The helicity violating Compton amplitudes $\mathcal{M}_{++}$ vanish in the forward limit and therefore matching against a cross section fixes the coefficient $c_{+-}$ uniquely, while leaving $c_{++}$ undetermined. In order to determine all parameters another observable is required. In \cite{Bautista:2021wfy,Bautista:2022wjf} the Compton amplitudes themselves were directly matched with black hole perturbation theory, while in \cite{Aoude:2023fdm} the absorption cross sections for individual spinning partial waves were used. It might also be possible distinguish the contributions by matching forward/backward asymmetry of the Compton differential cross section \cite{Holstein:2013kia}. Explicitly performing this matching for generic compact bodies is an interesting open problem; for the relevant case of a neutron star this information is ultimately a prediction of nuclear theory.  

Starting from the Compton amplitudes, we construct the 2-to-2 elastic amplitude by sewing generalized unitarity cuts. In our conventions the necessary 3-particle amplitudes are: for the scalar
\begin{equation}
\mathcal{M}_3\left(1_\psi,2_{\phi_1},3_{\overline{\phi}_1}\right) = Q_s\,,
\end{equation}
for electrodynamics
\begin{equation}
\mathcal{M}_3\left(1_\gamma^{+1},2_{\phi_1},3_{\overline{\phi}_1}\right) = -\sqrt{2}  Q_e \frac{[1|p_2 |\xi\rangle}{\langle 1 \xi\rangle}, \hspace{5mm} \mathcal{M}_3\left(1_\gamma^{-1},2_{\phi_1},3_{\overline{\phi}_1}\right) = \sqrt{2} Q_e \frac{\langle 1|p_2 |\xi]}{[1 \xi]},
\label{eq:3ptEM}
\end{equation}
and for gravity
\begin{equation}
\mathcal{M}_3\left(1_h^{+2},2_{\phi_1},3_{\overline{\phi}_1}\right) = -\frac{\kappa}{2}\frac{[1|p_2 |\xi\rangle^2}{\langle 1 \xi\rangle^2}, \hspace{5mm} \mathcal{M}_3\left(1_h^{-2},2_{\phi_1},3_{\overline{\phi}_1}\right) = -\frac{\kappa}{2} \frac{\langle 1|p_2 |\xi]^2}{[1 \xi]^2},
\label{eq:3ptGR}
\end{equation}
where $|\xi\rangle$ and $|\xi]$ are arbitrary auxiliary spinors. The case of the scalar is trivial and the resulting numerator, defined in (\ref{elasticgeneralform}), resulting from  the sewing procedure is 
\begin{equation}
    \mathcal{N}_4^{(\psi)}(s)=\frac{\kappa^2 Q_s^2 c^{(\psi)}}{2m_1}\,.
\end{equation}
For photons and gravitons, there are two different types of cuts contributing:
\newcommand{\cutPPNew}{\begin{tikzpicture}
        \begin{scope}
        \filldraw[fill=none,draw=none] (1,1) rectangle (5,-3);
        \draw[-,thick] (1,0)--(3,0);
        \draw[vector,double] (3,0)--(2,-2) ;
        \draw[fermion] (1,-2)--(2,-2);
        \draw[fermion] (2,-2)--(4,-2);
        \draw[-,thick] (3,0)--(5,0);
        \draw[vector,double] (3,0)--(4,-2)  ;
        \draw[fermion] (4,-2)--(5,-2);
        \filldraw[fill=gray!10] (3,0) circle (15pt) node {$\mathcal{M}_{c}$};
        \draw[dashed,thick,red] (2.8,-1.2)--(2,-0.9);
        \draw[dashed,thick,red] (4,-0.9)--(3.2,-1.2);
        \draw[dashed,thick,red] (3,-1.6)--(3,-2.4);
        \node at (2.2,-0.7) {$+$};
        \node at (3.8,-0.7) {$+$};
        \node at (2,-1.2) {$-$};
        \node at (4,-1.2) {$-$};
        \end{scope}
    \end{tikzpicture}}
\newcommand{\cutPMNew}{\begin{tikzpicture}
        \begin{scope}
        \filldraw[fill=none,draw=none] (1,1) rectangle (5,-3);
        \draw[-,thick] (1,0)--(3,0);
        \draw[vector,double] (3,0)--(2,-2) ;
        \draw[fermion] (1,-2)--(2,-2);
        \draw[fermion] (2,-2)--(4,-2);
        \draw[-,thick] (3,0)--(5,0);
        \draw[vector,double] (3,0)--(4,-2)  ;
        \draw[fermion] (4,-2)--(5,-2);
        \filldraw[fill=gray!10] (3,0) circle (15pt) node {$\mathcal{M}_{c}$};
        \draw[dashed,thick,red] (2.8,-1.2)--(2,-0.9);
        \draw[dashed,thick,red] (4,-0.9)--(3.2,-1.2);
        \draw[dashed,thick,red] (3,-1.6)--(3,-2.4);
        \node at (2.2,-0.7) {$-$};
        \node at (3.8,-0.7) {$+$};
        \node at (2,-1.2) {$+$};
        \node at (4,-1.2) {$-$};
        \end{scope}
    \end{tikzpicture}}
\begin{equation}
    \mathcal{C}_{++}=\vcenter{\hbox{\cutPPNew}}\,,\qquad \mathcal{C}_{-+}=\vcenter{\hbox{\cutPMNew}}\;\;\; ,
\end{equation}
the two remaining cuts $\mathcal{C}_{--}$ and $\mathcal{C}_{+-}$ are related by conjugation.
Using the Compton amplitudes in \eqref{eq:PhotonComptonPM} and \eqref{eq:PhotonComptonPP} as well as the three-point amplitudes in \eqref{eq:3ptEM} and making use of spinor identities listed in Appendix \ref{app:conventions}, for electrodynamics we calculate
\begin{align}
\mathcal{C}_{++}^{(\gamma)}(s)={}& -\frac{2\kappa^2  Q_e^2 c_{++}^{(\gamma)}m_1^2s|q|^2}{u_2\cdot \ell-s+\imath 0}+(u_2\to -u_2)\,,\\
\mathcal{C}_{-+}^{(\gamma)}(s)={}&-\frac{2\kappa^2  Q_e^2c_{-+}^{(\gamma)}m_1^2s}{u_2\cdot \ell-s+\imath 0}\frac{\left(y |q|^2 +2 \imath \varepsilon(\ell,q,u_1,u_2)\right)^2}{|q|^2}+(u_2\to -u_2)\,,
\end{align}
with $\varepsilon(\ell,q,u_1,u_2)\coloneqq\varepsilon_{\mu\nu\rho\sigma}\ell^\mu q^\nu u_1^\rho u_2^\sigma$. For gravity we find a simple \textit{double-copy}-like form
\begin{align}
\mathcal{C}_{++}^{(h)}(s)={}& -\frac{1}{4}\frac{\kappa^4 c_{++}^{(h)} m_1^4s|q|^4}{u_2\cdot \ell-s+\imath 0}+(u_2\to -u_2)\,,\\
\mathcal{C}_{-+}^{(h)}(s)={}&\frac{\kappa^4 c_{-+}^{(h)}  m_1^4s}{u_2\cdot \ell-s+\imath 0}\frac{\left(y |q|^2 +2\imath \varepsilon(\ell,q,u_1,u_2)\right)^4}{|q|^4}+(u_2\to -u_2)\,.
\end{align}
Summing over the four different cuts, yields the following numerators, up to pinch terms
\begin{align}
\mathcal{N}^{(\gamma)}_4(s)={}&2\kappa^2 Q_e^2m_1 \left\{\left[4s^2+(2y^2-1)|q|^2\right]c_{+-}^{(\gamma)}-|q|^2c_{++}^{(\gamma)}\right\}\,,\\
\mathcal{N}^{(h)}_4(s)={}&-\frac{\kappa^4 m_1^3}{4}\left\{\left[16s^4{+}8(4y^2-1)s^2|q|^2{+}(8y^4{-}8y^2{+}1)|q|^4\right]c_{+-}^{(h)}+|q|^4 c_{++}^{(h)}\right\}\,.
\end{align}
Starting from these expressions and using the mass-shift formula \eqref{massshiftformula} 
and the integrals \eqref{eq:MI20}--\eqref{eq:MI22},
it is straightforward to derive the leading-order mass-shifts for generic bodies
\begin{align}   
    \label{massshiftgeneral}
    \left(\Delta m_2\right)^{(\psi)} &= \frac{\pi G_\mathrm{N} Q_s^2 c^{(\psi)}}{32m_1^2 m_2|b|^3}\sqrt{y^2-1}\,\\
        \left(\Delta m_2\right)^{(\gamma)} &= \frac{9\pi G_\mathrm{N} Q_e^2}{32 m_2 |b|^5}\left[c_{+-}^{(\gamma)}\left(5y^2-1\right) -  4c_{++}^{(\gamma)}\right]\sqrt{y^2-1}\,,\\
        \left(\Delta m_2\right)^{(h)} &= \frac{225\pi^2 G_\mathrm{N}^2 m_1^2}{16m_2|b|^7}\left[c_{+-}^{(h)}\left(21y^4-14y^2+1\right)+8c_{++}^{(h)} \right]\sqrt{y^2-1}\,.
\end{align}
These expressions reduce to the Schwarzschild mass-shifts (\ref{massshiftscalar}), (\ref{massshiftphoton}) and (\ref{massshiftgraviton}) for the choice of dissipation numbers
\begin{equation}
    c^{(\psi)} = \frac{G_\mathrm{N} m_2^3}{\pi},\;\;\;\; c_{+-}^{(\gamma)} = \frac{G_\mathrm{N}^3 m_2^5}{3\pi}, \;\;\;\; c_{++}^{(\gamma)} = 0, \;\;\;\; c_{+-}^{(h)} = \frac{G_\mathrm{N}^5 m_2^7}{45\pi}, \;\;\;\; c_{++}^{(h)} = 0,
\end{equation}
where as discussed in Section \ref{subsec:crosssection}, the vanishing of the helicity violating amplitudes is a manifestation of (linearized) self-duality.

The general result (\ref{massshiftgeneral}) agrees with the contribution of the electric and magnetic components that can be read of from Eq.~(3.20) in \cite{Goldberger:2020wbx}. Let us point out the trivial fact that given that we have computed the absorption for models with scalar, vector and tensor exchange, we can also make predictions for more general theories including supergravity.\footnote{The contributions from fermionic exchanges are a bit more complicated and necessitate a process where the light particle transitions from a spin 0 to a higher-spin state.} 
In this context it is interesting to note that no matter which mediation fields are present in the theory, the helicity-conserving contribution of the graviton will always be dominant in the high-energy limit. This implies a universal high-energy behaviour for the mass-shift, which closely mirrors similar findings for the scattering angle discussed \cite{Bern:2020gjj,Parra-Martinez:2020dzs}.

\section{Discussion}
\label{sec:discussion}

The main results of this paper are the expressions (\ref{massshiftscalar}), (\ref{massshiftphoton}) and (\ref{massshiftgraviton}) for the change of rest mass of a Schwarzschild black hole scattering with a second compact body sourcing a massless scalar, electromagnetic or gravitational field. From these expressions we trivially obtain the 4-momentum impulse on either body using (\ref{deltaptodeltam}) and (\ref{deltap1deltap2}). The gravitational mass-shift (\ref{massshiftgraviton}) was previously calculated in \cite{Goldberger:2020wbx} using a Schwinger-Keldysh in-in path integral based on the worldline formalism introduced in \cite{Goldberger:2004jt,Goldberger:2005cd}; our result obtained using the scattering amplitudes based KMOC formalism \cite{Kosower:2018adc} is in complete agreement. As far as we are aware, the expressions for the impulse due to scalar (\ref{massshiftscalar}) and photon (\ref{massshiftphoton}) absorption are new. 

The case of massless scalar absorption is of some importance for recent comparisons between PM methods and (a simplified scalar toy-model of) the self-force expansion \cite{Barack:2022pde,Barack:2023oqp}. In the model considered in \cite{Barack:2022pde,Barack:2023oqp}, beginning at $\mathcal{O}\left(G_\mathrm{N}^2 Q_s^2\right)$, there is energy loss in the form of both radiation and horizon absorption. Combining the previous PM calculation of the two-loop radiative energy loss \cite{Barack:2023oqp} with the scalar absorption result of this paper (\ref{massshiftscalar}), we find excellent numerical agreement with the predicted energy loss from scalar self-force calculations at this order. Details of this comparison will be presented elsewhere. 

Beyond these specific results, in this paper we have developed an approach to incorporating horizon absorption based on familiar in-out scattering amplitudes. As explained in Section \ref{subsec:mass}, prior to integration, the non-trivial part of the calculation reduces to an application of standard unitarity methods for constructing integrands \cite{Bern:1994zx,Bern:1994cg,Bern:1995db,Bern:1997sc,Britto:2004nc}. Furthermore, as emphasized in Section \ref{subsec:Love}, the near-threshold expansion of the spectral integrals is most naturally treated as part of the expansion of the loop integrand in the soft region (\ref{soft}). Taken together, the approach described in this paper can be incorporated quite naturally into the existing workflow for scattering amplitudes based methods applied to the PM gravitational two-body problem.

An important application of these methods will be the incorporation of spin degrees-of-freedom and the calculation of absorptive effects in the scattering of Kerr black holes \cite{Porto:2007qi,Goldberger:2022rqf}. As observed in \cite{Chung:2018kqs,Guevara:2018wpp}, the static Kerr solution corresponds to a 3-particle on-shell amplitude previously identified as possessing uniquely soft UV behavior \cite{Arkani-Hamed:2017jhn}. There has been considerable effort in attempting to generalize this correspondence to 4-particle Compton amplitudes \cite{Chung:2018kqs,Guevara:2018wpp,Bautista:2019tdr,Aoude:2020onz,Saketh:2022wap,Chiodaroli:2021eug,Bern:2022kto,Cangemi:2022bew,Aoude:2022trd,Aoude:2022thd,Aoude:2023vdk,Bjerrum-Bohr:2023iey}. As emphasized recently \cite{Bautista:2021wfy,Bautista:2022wjf}, direct extraction of the Kerr Compton amplitude from the Teukolsky equation is complicated at higher-orders in spin by non-trivial absorption effects. We may hope that an intrinsically amplitudes based approach to horizon absorption will be useful in clarifying the situation.

While the discussion in this paper is presented having in mind the case of compact astronomical objects such as black holes and neutron stars, we note that the discussion should equally apply to elementary processes, in particular the low energy scattering of photons off heavy ions and pions. Compton scattering serves as an important probe in low energy QCD and the polarizabilities, e.g. of pions constitute key quantities to be computed from nuclear models (see e.g.~\cite{Schumacher:2005an,Dattoli:1977wg,Hecking:1981fsu,Weiner:1985ih}) and are measured e.g. at the COMPASS experiment at CERN \cite{COMPASS:2007rjf}. 

Finally, in this paper we have described the calculation of the \textit{leading} absorptive effects in two-body scattering. A significant simplification in this case was the fact that only 2-point functions of the $\mathcal{O}$ operators contributed, and therefore, as discussed in Section \ref{subsec:neutron}, our ignorance of the invisible sector could be parametrized in terms of a finite set of dissipation numbers arising from the near-threshold expansion of the spectral density. At higher-orders we would naively expect higher-point functions to contribute, and it is less clear how we can treat these quantities in a model-independent way. In \cite{Goldberger:2019sya} it was conjectured that the hidden sector dynamics is approximately Gaussian and that higher-point correlators factor into products of 2-point functions. To test this conjecture it may be useful to perform the EFT matching in the context of a toy model where the quantum mechanics of the black hole horizon is explicitly calculable \cite{Mertens:2022irh}. We leave this and other important open questions to future work.

\vspace{3mm}
\noindent \textbf{Acknowledgements}

\vspace{3mm}
We would like to thank Leor Barack, Zvi Bern, Enrico Herrmann, Dimitrios Kosmopoulos, Oliver Long and Chia-Hsien Shen for useful discussions and comments on the draft. CRTJ and MR are supported by the Department of Energy under Award Number DE-SC0009937, and gratefully acknowledge the continued support of the Mani L. Bhaumik Institute for Theoretical Physics. We would also like to acknowledge the hospitality of Nordita during the program \textit{Amplifying Gravity at All Scales} where part of this work was completed.

\appendix

\section{Conventions}
\label{app:conventions}

 We will use the mostly-minus metric convention, in $D$ spacetime dimensions
 \begin{equation}
    \eta_{\mu\nu} = \text{diag}(+1,\underbrace{-1,...,-1}_{D-1}).   
 \end{equation}
Loop and Fourier integrals are evaluated using dimensional regularization with $D=4-2\epsilon$. \\
\\
Our kinematic conventions for elastic scattering amplitudes will be to denote the incoming (outgoing) momenta $p_i$ ($p_i'$) using the 4-velocities $u_i$ and momentum transfer $q$ as
\begin{equation}
    p_1^\mu = m_1 u_1^\mu,\hspace{5mm} p_2^\mu = m_2 u_2^\mu, \hspace{5mm} {p_1'}^{\mu} = m_1 u_1^\mu + q^\mu, \hspace{5mm} {p_2'}^{\mu} = m_2 u_2^\mu - q^\mu.
\end{equation}
It is also convenient to introduce the \textit{dual} vectors
\begin{equation}
    \label{dualvectors}
    \check{u}_1^\mu \coloneqq\frac{u_1^\mu - y u_2^\mu}{1-y^2}, \hspace{10mm} \check{u}_2^\mu \coloneqq\frac{u_2^\mu - y u_1^\mu}{1-y^2}, \hspace{10mm} y \coloneqq u_1 \cdot u_2,
\end{equation}
defined to satisfy $u_i \cdot \check{u}_j = \delta_{ij}$.\\
\\
Following \cite{Kosower:2018adc} we absorb many factors of $2\pi$ by defining 
 \begin{equation}
     \hat{\text{d}} x \coloneqq\frac{\text{d} x}{2\pi}, \hspace{10mm} \hat{\delta}\left(x\right) \coloneqq2\pi \delta\left(x\right).
 \end{equation}
 Our scattering conventions mostly follow the conventions of Peskin-Schroeder \cite{Peskin:1995ev}. Scattering amplitudes are related to the formal S-matrix operator by
 \begin{equation}
     \langle \text{out}|T|\text{in}\rangle \coloneqq\hat{\delta}^{(D)}\left(p_{\text{out}}-p_{\text{in}}\right)\mathcal{M}\left(\text{in}\rightarrow \text{out}\right),
 \end{equation}
 where $S \coloneqq1+\imath T$. All of the amplitudes needed in the paper can be calculated using the Feynman rules enumerated in Appendix \ref{app:Feynman}. We use the standard phase convention, these rules directly applied calculate $\imath\mathcal{M}$.\\
\\
Our spinor-helicity conventions mostly follow Elvang-Huang \cite{Elvang:2013cua}, modified to align with the choice of metric signature. All Pauli matrix identities and definitions are chosen to align with \cite{Dreiner:2008tw}. We choose $|p]_\alpha$ and $\langle p|_{\dot{\alpha}}$ to carry (outgoing) helicity weights $+1/2$ and $-1/2$ respectively and for real momenta $\left(|p]_\alpha\right)^* = \langle p|_{\dot{\alpha}}$. We use the convention in which the helicity spinors and bispinor momenta are related as
\begin{equation}
    p_\mu \sigma^\mu_{\alpha\dot{\alpha}} \coloneqq p_{\alpha\dot{\alpha}} \coloneqq -|p]_\alpha \langle p|_{\dot{\alpha}}.
\end{equation}
A consistent choice of (outgoing) spin-1 polarization vectors is given by
\begin{equation}
    \varepsilon_+^{*\mu}\left(p\right) \coloneqq\frac{[p|\sigma^\mu |\xi\rangle}{\sqrt{2}\langle p \xi\rangle}, \hspace{10mm} \varepsilon_-^{*\mu}\left(p\right) \coloneqq-\frac{\langle p|\overline{\sigma}^\mu |\xi]}{\sqrt{2}[p \xi]},
\end{equation}
where $|\xi\rangle$ and $|\xi]$ are arbitrary auxiliary vectors; this choice is consistent with the property $\left(\varepsilon_+^\mu\right)^* = \varepsilon_-^\mu$ as well as the normalization and completeness conditions
\begin{equation}
\varepsilon_\lambda^*\left(p\right)\cdot \varepsilon_{\lambda^\prime}\left(p\right)=-\delta_{\lambda\lambda^\prime},\,\quad \hspace{10mm} \sum_{\lambda=\pm}(\varepsilon_\lambda^\mu\left(p\right))^* \varepsilon_\lambda^\nu\left(p\right) = -\eta^{\mu\nu} + \frac{p^\mu \xi^\nu+p^\nu \xi^\mu}{p\cdot\xi}.
\end{equation}

\section{Duality-Organized Operator Basis \label{sec:ExplicitLagr}}
By adopting an on-shell approach as we have done in Section \ref{subsec:neutron}, we never have to introduce an off-shell effective action. However for the convenience of the reader and in order to facilitate comparison with the literature, we explicitly give the operators that when added to the minimal action give the corresponding Compton amplitudes.
The photon Compton amplitudes in \eqref{eq:PhotonComptonPM} and \eqref{eq:PhotonComptonPP} can be obtained from
\begin{align}
    S^{(\gamma)}_{\text{portal}}={}& 2\kappa\int\mathrm{d}^4x\,\left[   \frac{\partial_\mu\phi_2}{m_2}F_{\mathrm{SD}}^{\mu\nu}\mathcal{O}_\nu^{\mathrm{SD}}+ \frac{\partial_\mu\phi_2}{m_2}F_{\mathrm{ASD}}^{\mu\nu}\mathcal{O}_\nu^{\mathrm{ASD}}\right]\,,
\end{align}
where the self-dual and anti-self-dual field strengths are given by
\begin{equation}
    F_{\mathrm{SD}}^{\mu\nu}=\frac{1}{2}\left(F^{\mu\nu}+\frac{\imath}{2}\epsilon^{\mu\nu}_{\phantom{\mu\nu}\alpha\beta}F^{\alpha\beta}\right)\,,  \quad  F_{\mathrm{ASD}}^{\mu\nu}=\frac{1}{2}\left(F^{\mu\nu}-\frac{\imath}{2}\epsilon^{\mu\nu}_{\phantom{\mu\nu}\alpha\beta}F^{\alpha\beta}\right)\,.
\end{equation}
The fact that this basis is particularly amplitude friendly is well-known and frequently used in the context of EFT (see e.g. \cite{Dixon:2004za}). The correlation functions are given by
\begin{align}
    \langle \mathcal{O}_{\mathrm{SD}}^{\mu}(x) \mathcal{O}_{\mathrm{ASD}}^{\nu}(0)\rangle ={}&-i\int \mathrm{d}^D p\, e^{\imath p\cdot x}\int_{m_2^2}^\infty\mathrm{d}\mu^2\frac{c_{+-}^{(\gamma)}(\mu^2-m_2^2)}{p^2-\mu^2+\imath 0}\Pi_1^{\mu\nu}(p)+\dots\\
    \langle \mathcal{O}_{\mathrm{SD}}^{\mu}(x) \mathcal{O}_{\mathrm{SD}}^{\nu}(0)\rangle ={}&-i\int \mathrm{d}^D p\, e^{\imath p\cdot x}\int_{m_2^2}^\infty\mathrm{d}\mu^2\frac{c_{++}^{(\gamma)}(\mu^2-m_2^2)}{p^2-\mu^2+\imath 0}\Pi_1^{\mu\nu}(p)+\dots
\end{align}
The mixed other correlators are related by parity. In this we have introduced the projector
\begin{equation}
    \Pi_1^{\mu\nu}(p)=\eta^{\mu\nu}-\frac{p^\mu p^\nu}{p^2}\,.
\end{equation}
Where as before the ellipsis denote higher-order-in-$G_\mathrm{N}$ and quantum terms. The operators for the gravitational case look very similar and are explicitly 
\begin{align}
    S_{\text{portal}}^{(h)} ={} &\int \text{d}^4 x \left[ \frac{\partial^{\alpha}\partial^{\beta}\phi_2}{m_2^2} C_{\mu\alpha\nu\beta}^{\mathrm{SD}}\mathcal{O}_{\mathrm{SD}}^{\mu\nu}+\frac{\partial^{\alpha}\partial^{\beta}\phi_2}{m_2^2} C_{\mu\alpha\nu\beta}^{\mathrm{ASD}}\mathcal{O}_{\mathrm{ASD}}^{\mu\nu}\right]\,.
\end{align}
The self-and anti-self-dual projections of the Weyl tensor are 
\begin{equation}
    C_{\mu\nu\rho\sigma}^{\mathrm{SD}}=\frac{1}{2}\left( C_{\mu\nu\rho\sigma}+\frac{\imath}{2}\epsilon_{\mu\nu\alpha\beta}C^{\alpha\beta}_{\phantom{\mu\nu}\rho\sigma}\right)\,,\quad C_{\mu\nu\rho\sigma}^{\mathrm{ASD}}=\frac{1}{2}\left( C_{\mu\nu\rho\sigma}-\frac{\imath}{2}\epsilon_{\mu\nu\alpha\beta}C^{\alpha\beta}_{\phantom{\mu\nu}\rho\sigma}\right)\,.
\end{equation}
The correlators are
\begin{align}
    \langle \mathcal{O}_{\mathrm{SD}}^{\mu\nu}(x) \mathcal{O}_{\mathrm{ASD}}^{\alpha \beta}(0)\rangle ={}&-\imath\int \mathrm{d}^D p\, e^{\imath p\cdot x}\int_{m_2^2}^\infty\mathrm{d}\mu^2\frac{c_{+-}^{(h)}(\mu^2-m_2^2)}{p^2-\mu^2+\imath 0}\Pi_2^{\mu\nu\alpha\beta}(p)+\dots\,,\\
    \langle \mathcal{O}_{\mathrm{SD}}^{\mu\nu}(x) \mathcal{O}_{\mathrm{SD}}^{\alpha\beta}(0)\rangle ={}&-\imath\int \mathrm{d}^D p\, e^{\imath p\cdot x}\int_{m_2^2}^\infty\mathrm{d}\mu^2\frac{c_{++}^{(h)}(\mu^2-m_2^2)}{p^2-\mu^2+\imath 0}\Pi_2^{\mu\nu\alpha\beta}(p)+\dots\,,
\end{align}
once again the remaining correlators are related by parity; the projector is given by\footnote{
In practice the correlator will always contract into an conserved current, so we can replace $\Pi_1^{\mu\nu}\to \eta^{\mu\nu}$ and $\Pi^{\mu\nu\alpha\beta}_2\to\eta^{\mu\alpha}\eta^{\nu\beta}$.}
\begin{equation}
    \Pi^{\mu\nu\alpha\beta}_2(p)=\frac{1}{2}\Pi_1^{\mu\alpha}(p)\Pi_1^{\nu\beta}(p)+\frac{1}{2}\Pi_1^{\mu\beta}(p)\Pi_1^{\nu\alpha}(p)-\frac{1}{D-1}\Pi_1^{\mu\nu}(p)\Pi_1^{\alpha\beta}(p)\,.
\end{equation}

\section{Weighted Cuts}
\label{app:cuts}
 
 It is convenient to define a diagrammatic \textit{weighted} cut as a generalization of the standard Cutkosky cutting rules \cite{Cutkosky}:
 \begin{center}
    \begin{tikzpicture}
        \draw (-1,1)--(0,0);
        \draw (-1,-1)--(0,0);
        \node at (-1,0.1) {\LARGE$\vdots$};
        \draw (3,0)--(4,1);
        \draw (3,0)--(4,-1);
        \node at (4,0.1) {\LARGE$\vdots$};
        \draw (0,0) arc (150:30:1.7);
        \draw (0,0) arc (-150:-30:1.7);
        \draw[dashed,thick,red] (1.5,1.2)--(1.5,0.4);
        \draw[dashed,thick,red] (1.5,-0.4)--(1.5,-1.2);
        \draw[->] (0.5,0.8)--(0.9,1);
        \draw[->] (0.5,-0.8)--(0.9,-1);
        \node at (0.7,1.3) {$l_1$};
        \node at (0.7,-1.3) {$l_2$};
        \node at (1.5,0) {\color{red}$F(l_i)$\color{black}};
        \filldraw[fill=gray] (0,0) circle (15pt);
        \filldraw[fill=gray] (3,0) circle (15pt);
        \node at (5,0) {$\coloneqq$};
        \node at (9,-0.1) {
        \begin{minipage}{10cm}
            \begin{equation}
                \imath\mathcal{M}\biggr\vert_{\frac{1}{[d_1+\imath 0][d_2+\imath 0]}\rightarrow (-\imath)^2 F(l_i)\hat{\delta}^{(+)}\left(d_1\right)\hat{\delta}^{(+)}\left(d_2\right) },
            \end{equation}
        \end{minipage}
        };
        \node at (15,0) {};
    \end{tikzpicture}
\end{center}
where $d_i \coloneqq l_i^2 -m_i^2$ and $F(l_i)$ is an arbitrary function we will refer to as the \textit{weight} of the cut. Choosing $F(l_i)=1$ gives an ordinary Cutkosky cut that calculates the imaginary part of the amplitude \cite{Cutkosky}:
 \begin{center}
    \begin{tikzpicture}
        \draw (-1,1)--(0,0);
        \draw (-1,-1)--(0,0);
        \node at (-1,0.1) {\LARGE$\vdots$};
        \draw (3,0)--(4,1);
        \draw (3,0)--(4,-1);
        \node at (4,0.1) {\LARGE$\vdots$};
        \draw (0,0) arc (150:30:1.7);
        \draw (0,0) arc (-150:-30:1.7);
        \draw[dashed,thick,red] (1.5,1.2)--(1.5,0.4);
        \draw[dashed,thick,red] (1.5,-0.4)--(1.5,-1.2);
        \draw[->] (0.5,0.8)--(0.9,1);
        \draw[->] (0.5,-0.8)--(0.9,-1);
        \node at (0.7,1.3) {$l_1$};
        \node at (0.7,-1.3) {$l_2$};
        \node at (1.5,0) {\color{red}$1$\color{black}};
        \filldraw[fill=gray] (0,0) circle (15pt);
        \filldraw[fill=gray] (3,0) circle (15pt);
        \node at (8.5,0.1) {
        \begin{minipage}{11cm}
            \begin{equation}
                = \hspace{9mm}\text{Disc}\left(\imath\mathcal{M}\right) \hspace{9mm} = \hspace{9mm} -2\;\text{Im}\mathcal{M}.
            \end{equation}
        \end{minipage}
        };
        \node at (10,0) {};
    \end{tikzpicture}
\end{center}

\section{Feynman Rules}
\label{app:Feynman}

\subsection*{Scalar Model}

\begin{center}
    \begin{tikzpicture}
        \draw[fermion] (0,0)--(2,0);
        \node at (1,0.4) {$k$};
        \node at (5,0.2) {
        \begin{minipage}{3cm}
            \begin{equation*}
                = \hspace{12mm}\frac{\imath}{k^2-m_1^2+\imath 0}.
            \end{equation*}
        \end{minipage}
        };
    \end{tikzpicture}
\end{center}
\begin{center}
    \begin{tikzpicture}
        \draw[dashed] (0,0)--(2,0);
        \draw[->] (0.5,0.3)--(1.5,0.3);
        \node at (1,0.6) {$k$};
        \node at (5,0.2) {
        \begin{minipage}{3cm}
            \begin{equation*}
                = \hspace{12mm}\frac{\imath}{k^2+\imath 0}.
            \end{equation*}
        \end{minipage}
        };
    \end{tikzpicture}
\end{center}
\begin{center}
    \begin{tikzpicture}
        \draw[dashed] (-1.5,0)--(0,0);
        \draw[fermion] (0,0)--(0.75,1.3);
        \draw[fermionbar] (0,0)--(0.75,-1.3);
        \node at (3.5,0.2) {
        \begin{minipage}{3cm}
            \begin{equation*}
                = \hspace{12mm}\imath Q_s.
            \end{equation*}
        \end{minipage}
        };
    \end{tikzpicture}
\end{center}
\begin{center}
    \begin{tikzpicture}
        \draw[-,double] (0,0)--(2,0);
        \draw[->] (0.5,0.3)--(1.5,0.3);
        \node at (1,0.6) {$k$};
        \node at (5,0.2) {
        \begin{minipage}{3cm}
            \begin{equation*}
                = \hspace{12mm}\int_0^\infty \text{d}\mu^2\frac{\imath\rho_0(\mu^2)}{k^2-\mu^2+\imath 0}.
            \end{equation*}
        \end{minipage}
        };
    \end{tikzpicture}
\end{center}
\begin{center}
    \begin{tikzpicture}
        \draw[dashed] (-1.5,0)--(0,0);
        \draw[-,double] (0,0)--(0.75,1.3);
        \draw[-,thick] (0,0)--(0.75,-1.3);
        \filldraw (0,0) circle (3pt);
        \node at (2.5,0.2) {
        \begin{minipage}{3cm}
            \begin{equation*}
                = \hspace{12mm}\imath\kappa.
            \end{equation*}
        \end{minipage}
        };
    \end{tikzpicture}
\end{center}

\subsection*{Scalar QED}

\begin{center}
    \begin{tikzpicture}
        \draw[vector] (0,0)--(2,0);
        \draw[->] (0.5,0.3)--(1.5,0.3);
        \node at (-0.2,0) {$\mu$};
        \node at (2.2,0) {$\nu$};
        \node at (1,0.6) {$k$};
        \node at (5,0.2) {
        \begin{minipage}{3cm}
            \begin{equation*}
                = \hspace{12mm}\frac{-\imath\eta^{\mu\nu}}{k^2+\imath 0}.
            \end{equation*}
        \end{minipage}
        };
    \end{tikzpicture}
\end{center}
\begin{center}
    \begin{tikzpicture}
        \draw[vector] (-1.5,0)--(0,0);
        \draw[fermion] (0,0)--(0.75,1.3);
        \draw[fermionbar] (0,0)--(0.75,-1.3);
        \node at (0.7,-0.7) {$p$};
        \node at (0.7,0.7) {$p'$};
        \node at (-1.7,0) {$\mu$};
        \node at (3.5,0.2) {
        \begin{minipage}{3cm}
            \begin{equation*}
                = \hspace{12mm}-\imath Q_e\left(p^\mu + p'^\mu\right).
            \end{equation*}
        \end{minipage}
        };
    \end{tikzpicture}
\end{center}
\begin{center}
    \begin{tikzpicture}
        \draw[-,double] (0,0)--(2,0);
        \draw[->] (0.5,0.3)--(1.5,0.3);
        \node at (1,0.6) {$k$};
        \node at (-0.5,0) {$\mu\nu$};
        \node at (2.5,0) {$\rho\sigma$};
        \node at (5,0.2) {
        \begin{minipage}{3cm}
            \begin{equation*}
                = \hspace{12mm}\int_0^\infty \text{d}\mu^2\rho_1\left(\mu^2\right)\frac{\imath\hat{\Pi}_1^{\mu\nu\rho\sigma}(k)}{k^2-\mu^2+\imath 0},
            \end{equation*}
        \end{minipage}
        };
    \end{tikzpicture}
\end{center}
where $\hat{\Pi}_1$ is given in (\ref{hatPi1}). 
\begin{center}
    \begin{tikzpicture}
        \draw[vector] (-1.5,0)--(0,0);
        \draw[-,double] (0,0)--(0.75,1.3);
        \draw[thick] (0,0)--(0.75,-1.3);
        \node at (-1.7,0) {$\mu$};
        \node at (0.9,1.5) {$\nu\rho$};
        \draw[->] (-0.3,0.3)--(-1.2,0.3);
        \node at (-0.75,0.5) {$p$};
        \filldraw (0,0) circle (3pt);
        \node at (3.5,0.2) {
        \begin{minipage}{3cm}
            \begin{equation*}
                = \hspace{12mm}-\kappa\left(\eta^{\mu\nu}p^\rho-\eta^{\mu\rho}p^\nu\right).
            \end{equation*}
        \end{minipage}
        };
    \end{tikzpicture}
\end{center}

\subsection*{General Relativity}

\begin{center}
    \begin{tikzpicture}
        \draw[vector,double] (0,0)--(2,0);
        \draw[->] (0.5,0.3)--(1.5,0.3);
        \node at (-0.5,0) {$\mu\nu$};
        \node at (2.5,0) {$\rho\sigma$};
        \node at (1,0.6) {$k$};
        \node at (5,0.2) {
        \begin{minipage}{3cm}
            \begin{equation*}
                = \hspace{12mm}\frac{\frac{\imath}{2}\left(\eta^{\mu\rho}\eta^{\nu\sigma}+\eta^{\mu\sigma} \eta^{\nu\rho} - 
 \eta^{\mu\nu} \eta^{\rho\sigma}\right)}{k^2+\imath 0}.
            \end{equation*}
        \end{minipage}
        };
        \node at (11.5,0) {};
    \end{tikzpicture}
\end{center}
\begin{center}
    \begin{tikzpicture}
        \draw[vector,double] (-1.5,0)--(0,0);
        \draw[fermion] (0,0)--(0.75,1.3);
        \draw[fermionbar] (0,0)--(0.75,-1.3);
        \node at (0.7,-0.7) {$p$};
        \node at (0.7,0.7) {$p'$};
        \node at (-1.9,0) {$\mu\nu$};
        \node at (3.5,0.2) {
        \begin{minipage}{3cm}
            \begin{equation*}
                = \hspace{12mm}\frac{\imath\kappa}{2} \left(\eta^{\mu\nu} ((p\cdot p') - m_1^2) - 
   p^\mu {p'}^{\nu} - p^{\nu} {p'}^{\mu}\right).
            \end{equation*}
        \end{minipage}
        };
        \node at (10,0) {};
    \end{tikzpicture}
\end{center}
\begin{center}
    \begin{tikzpicture}
        \draw[-,double] (0,0)--(2,0);
        \draw[->] (0.5,0.3)--(1.5,0.3);
        \node at (1,0.6) {$k$};
        \node at (-1,0) {$\mu_1\mu_2\mu_3\mu_4$};
        \node at (3,0) {$\nu_1\nu_2\nu_3\nu_4$};
        \node at (6,0.2) {
        \begin{minipage}{3cm}
            \begin{equation*}
                = \hspace{12mm}\int_0^\infty \text{d}\mu^2\rho_2\left(\mu^2\right)\frac{\imath\hat{\Pi}_2^{\mu_1\mu_2\mu_3\mu_4\nu_1\nu_2\nu_3\nu_4}(k)}{k^2-\mu^2+\imath 0},
            \end{equation*}
        \end{minipage}
        };
        \node at (12,0) {};
    \end{tikzpicture}
\end{center}
where $\hat{\Pi}_2 \coloneqq\mathcal{P} \cdot \Sigma(k)$, with
    \begin{align}
    \label{gravitonnumerator1}
    &\mathcal{P}^{\mu_1 \mu_2 \mu_3 \mu_4 \nu_1 \nu_2 \nu_3 \nu_4}_{\rho_1 \rho_2 \rho_3\rho_4 \sigma_1 \sigma_2 \sigma_3 \sigma_4} \nonumber\\
    &\coloneqq\frac{1}{8}\left[\left(\delta^{[\mu_1}_{\rho_1} \delta^{\mu_2]}_{\rho_2} \delta^{[\mu_3}_{\rho_3} \delta^{\mu_4]}_{\rho_4}+\delta^{[\mu_3}_{\rho_1} \delta^{\mu_4]}_{\rho_2} \delta^{[\mu_1}_{\rho_3} \delta^{\mu_2]}_{\rho_4}\right)\left(\delta^{[\nu_1}_{\sigma_1} \delta^{\nu_2]}_{\sigma_2} \delta^{[\nu_3}_{\sigma_3} \delta^{\nu_4]}_{\sigma_4}+\delta^{[\nu_3}_{\sigma_1} \delta^{\nu_4]}_{\sigma_2} \delta^{[\nu_1}_{\sigma_3} \delta^{\nu_2]}_{\sigma_4}\right)\right. \nonumber\\
    &\hspace{15mm}+\left. \left(\delta^{[\nu_1}_{\rho_1} \delta^{\nu_2]}_{\rho_2} \delta^{[\nu_3}_{\rho_3} \delta^{\nu_4]}_{\rho_4}+\delta^{[\nu_3}_{\rho_1} \delta^{\nu_4]}_{\rho_2} \delta^{[\nu_1}_{\rho_3} \delta^{\nu_2]}_{\rho_4}\right)\left(\delta^{[\mu_1}_{\sigma_1} \delta^{\mu_2]}_{\sigma_2} \delta^{[\mu_3}_{\sigma_3} \delta^{\mu_4]}_{\sigma_4}+\delta^{[\mu_3}_{\sigma_1} \delta^{\mu_4]}_{\sigma_2} \delta^{[\mu_1}_{\sigma_3} \delta^{\mu_2]}_{\sigma_4}\right)\right],
\end{align}
and 
\begin{align}
    \label{gravitonnumerator2}
   &\Sigma^{\mu_1\mu_2\mu_3\mu_4\nu_1\nu_2\nu_3\nu_4}(k) \nonumber\\
   &= \frac{8}{3} \left(2 \eta^{\mu_1\nu_4} \eta^{\mu_2\nu_3}
   \eta^{\mu_3\nu_2} \eta^{\mu_4\nu_1}+2 \eta^{\mu_1\nu_4}
   \eta^{\mu_2\nu_2} \eta^{\mu_3\nu_3} \eta^{\mu_4\nu_1}-\eta^{\mu_1\mu_4} \eta^{\mu_2\mu_3} \eta^{\nu_1\nu_4}
   \eta^{\nu_2\nu_3}\right)\nonumber\\
   &\hspace{5mm}-\frac{32}{ k^{2}} \left(2 k^{\mu_1} k^{\mu_3} \eta^{\mu_2\nu_4} \eta^{\mu_4\nu_2} \eta^{\nu_1\nu_3}+2
   k^{\mu_1} k^{\nu_1} \eta^{\mu_2\nu_4} \eta^{\mu_3\nu_3} \eta^{\mu_4\nu_2}-2 k^{\mu_3} k^{\nu_1}
   \eta^{\mu_1\nu_4} \eta^{\mu_2\nu_3} \eta^{\mu_4\nu_2}\right.\nonumber\\
   &\hspace{20mm}\left.+k^{\mu_1} k^{\mu_3} \eta^{\mu_2\mu_4} \eta^{\nu_1\nu_4} \eta^{\nu_2\nu_3}+2 k^{\mu_1} k^{\nu_1}
   \eta^{\mu_2\mu_4} \eta^{\mu_3\nu_4} \eta^{\nu_2\nu_3}\right)\nonumber\\
   &\hspace{5mm}+\frac{128
   k^{\mu_1} k^{\mu_3} k^{\nu_1} k^{\nu_3}}{(k^2)^2} \left(2
   \eta^{\mu_2\nu_4} \eta^{\mu_4\nu_2}-\eta^{\mu_2\mu_4}
   \eta^{\nu_2\nu_4}\right).
\end{align}
\begin{center}
    \begin{tikzpicture}
        \draw[vector,double] (-1.5,0)--(0,0);
        \draw[-,double] (0,0)--(0.75,1.3);
        \draw[thick] (0,0)--(0.75,-1.3);
        \node at (-1.9,0) {$\alpha\beta$};
        \node at (0.9,1.5) {$\mu\nu\rho\sigma$};
        \draw[->] (-0.3,0.3)--(-1.2,0.3);
        \node at (-0.75,0.5) {$p$};
        \filldraw (0,0) circle (3pt);
        \node at (8,-0.4) {
        \begin{minipage}{3cm}
            \begin{align*}
                = \hspace{8mm}&-\frac{\imath\kappa}{4}\left[\eta_{\mu\alpha}\eta_{\sigma\beta}p_\nu p_\rho - \eta_{\nu\alpha}\eta_{\sigma\beta} p_\mu p_\rho+\eta_{\nu\alpha}\eta_{\rho\beta}p_\mu p_\sigma\right. \nonumber\\
                &\hspace{10mm}\left.- \eta_{\mu\alpha}\eta_{\rho\beta}p_\nu p_\sigma+\eta_{\mu\beta}\eta_{\sigma\alpha}p_\nu p_\rho - \eta_{\nu\beta}\eta_{\sigma\alpha} p_\mu p_\rho\right. \nonumber\\
                &\hspace{10mm}\left.+\eta_{\nu\beta}\eta_{\rho\alpha}p_\mu p_\sigma - \eta_{\mu\beta}\eta_{\rho\alpha}p_\nu p_\sigma\right].
            \end{align*}
        \end{minipage}
        };
    \end{tikzpicture}
\end{center}

\section{Master Integrals}
\label{app:integrals}

To calculate the mass-shift (\ref{massshiftformula}) there are three integrals to evaluate, which can in principle be performed in any order.
The integrals in question, in general take the form 
\begin{equation}
    I_{k,n}=\int_0^\infty\mathrm{d}s\,s^k\int\hat{\mathrm{d}}^{4}q\, \hat{\delta}( u_2\cdot q)\hat{\delta}(u_1\cdot q)
    e^{\imath q\cdot b}|q|^{2n}\int\hat{\mathrm{d}}^{4}\ell\frac{\hat{\delta}( u_2\cdot \ell -s)\hat{\delta}(u_1\cdot \ell)}{\ell^2(\ell+q)^2}\,,
\end{equation}
where we set $D=4$ since all integrals that are required in the main text are finite. Furthermore we will only need the case where $k=2r$,  $r\in \mathds{Z}_{\geq 0}$.
The approach we will take is to first evaluate the spectral integral using
\begin{equation}
    \int_0^\infty \text{d}s\; s^{2r} \hat{\delta}\left(u_2\cdot \ell-s\right) = 2\pi(u_2\cdot \ell)^{2r} \theta\left(u_2\cdot \ell\right)\,.
\end{equation}
By making a change of variables $\ell\rightarrow -\ell-q$ and using the trivial identity $\theta(x)+\theta(-x)=1$, the resulting loop integrals can then be simplified using 
\begin{equation}
    \int \hat{\text{d}}^4 \ell \left[(u_2\cdot \ell)^{2r} \theta\left(u_2\cdot l\right)\right]\frac{\hat{\delta}\left(u_1\cdot \ell\right)}{\ell^2(\ell+q)^2} = \frac{1}{2} \int \hat{\text{d}}^4 \ell \frac{(u_2\cdot \ell)^{2r}\hat{\delta}\left(u_1\cdot \ell\right)}{\ell^2(\ell+q)^2}.
\end{equation}
Next we need scalar loop integrals of the form 
\begin{equation}
    \label{masterscalar}
    \int \hat{\text{d}}^{4} \ell \frac{(u_2\cdot \ell)^{2r}\hat{\delta}\left(u_1\cdot \ell\right)}{\ell^2(\ell+q)^2} \;\; = \;\; \frac{(-1)^r(2r)!}{2^{4r+3}(r!)^2}\left(y^2-1\right)^{r}\left(-q^2\right)^{\frac{2r-1}{2}}, \hspace{5mm} r\in \mathds{Z}_{\geq 0}.
\end{equation}
This formula is derived by specializing to the rest frame of particle-1
\begin{equation}
    u_1^\mu = (1,0,\vect{0}), \hspace{10mm} u_2^\mu = (y,\sqrt{y^2-1},\vect{0}), \hspace{10mm} q^\mu = (0,0,\vect{q}),\label{eq:restframe} 
\end{equation}
where
\begin{equation}
    y = u_1\cdot u_2,\hspace{10mm} |\vect{q}| = \left(-q^2\right)^{1/2},
\end{equation}
giving 
\begin{align}
     \int \hat{\text{d}}^4 \ell \;\frac{\left(u_2\cdot \ell\right)^{2r}\hat{\delta}\left(u_1\cdot \ell\right)}{\ell^2(\ell+q)^2} &=  \int \hat{\text{d}}\ell_0 \hat{\text{d}} \ell_1\hat{\text{d}}^{2} \vect{\ell} \;\frac{\left(y \ell_0 - \sqrt{y^2-1}\ell_1\right)^{2r}\hat{\delta}\left(\ell_0\right)}{[\ell_0^2-\ell_1^2 -\vect{\ell}^2][\ell_0^2-\ell_1^2-(\vect{\ell}+\vect{q})^2]}\nonumber\\
     &=  \frac{\left(y^2-1\right)^{r}}{\pi}\int_0^\infty  \text{d} \ell_1\;\ell_1^{2r}\int \hat{\text{d}}^{2} \vect{\ell} \;\frac{1}{[\vect{\ell}^2+\ell_1^2][(\vect{\ell}+\vect{q})^2+\ell_1^2]}\nonumber\\
    &=  \frac{\left(y^2-1\right)^{r}}{\pi}\int_0^\infty  \text{d} \ell_1\;\ell_1^{2r}\left[\frac{\text{arcsinh}\left(\frac{|\vect{q}|}{2\ell_1}\right)}{\pi|\vect{q}|\sqrt{\vect{q}^2+4 \ell_1^2}}\right] \nonumber\\
    &= \left(y^2-1\right)^{r}\left(-q^2\right)^{\frac{2r-1}{2}}\times \frac{1}{\pi^2}\int_0^\infty  \text{d} x\; x^{2r} \left[\frac{\text{arcsinh}\left(\frac{1}{2x}\right)}{\sqrt{1+4x^2}}\right].
\end{align}
The remaining integral is just a number, but diverges for $r\in \mathds{Z}_{>0}$; we evaluate it for $r\in \mathds{R}$ in the range $-1/2<r<1/2$ and analytically continue the result. 
Finally we take the Fourier transform to impact parameter space. The derivation of the general result is well-known, for example Appendix A of \cite{Herrmann:2021tct}, and is repeated here for convenience 
\begin{align}
    \label{masterFourier}
    &\int \hat{\text{d}}^4 q \;\hat{\delta}\left(u_1\cdot q\right) \hat{\delta}\left(u_2\cdot q\right)  e^{\imath q\cdot b} \left(-q^2\right)^{-\alpha} = \frac{1}{\sqrt{y^2-1}} \frac{\Gamma\left(1-\alpha\right)}{2^{2\alpha}\pi\Gamma(\alpha)} |b|^{2\alpha-2},
\end{align}
where by assumption $u_1\cdot b = u_2 \cdot b = 0$.
For completeness, the cases needed in the main text
\begin{align}
    I_{2,0}={}&\frac{1}{128}\frac{\sqrt{y^2-1}}{|b|^3}\,,\label{eq:MI20}\\
    I_{4,0}={}&\frac{27}{2048 }\frac{\left(y^2-1\right)^{3/2}}{|b|^5}\,,\\
    I_{2,1}={}&-\frac{9}{128}\frac{\sqrt{y^2-1}}{|b|^5}\,,\\
    I_{6,0}={}&\frac{1125}{16384}\frac{\left(y^2-1\right)^{5/2}}{|b|^7}\,,\\
    I_{4,1}={}&-\frac{675}{2048}\frac{\left(y^2-1\right)^{3/2}}{|b|^7}\,,\\
    I_{2,2}={}&\frac{225 }{128 }\frac{\sqrt{y^2-1}}{|b|^7}\label{eq:MI22}\,.
\end{align}
The integral $I_{2,0}$ is needed in the scalar computation, $I_{4,0},I_{2,1}$ are needed for the photon and $I_{6,0},I_{4,1},I_{2,2}$ are needed for the graviton.

Let us also give an alternative derivation. First note that we can set $n=0$, all other integrals are obtained through the usual trick of taking derivatives with respect of $b$. Specifying to the rest frame of particle-1 in (\ref{eq:restframe}) and performing the $\ell_0$ and $\ell_1$ integrations which sets $\ell_0=0$ and $\ell_1=-\frac{s}{\sqrt{y^2-1}}$, we find

\begin{align}
    	I_{k,0}=\frac{1}{(y^2-1)}\int_0^\infty\mathrm{d}s\,s^k{}& \int\hat{\mathrm{d}}^{2}\vect{q}\,e^{-\imath \vect{q}\cdot \vect{b}}
     \int \frac{\hat{\mathrm{d}}^{2}\vect{\ell} }{[\vect{\ell}^2+s^2/(y^2-1)][(\vect{\ell}+\vect{q})^2+s^2/(y^2-1)]}.
\end{align}
We notice that this is the Fourier transform of a convolution, such that
\begin{align}
    	I_{k,0}={}&\frac{1}{(y^2-1)}	\int_0^\infty\mathrm{d}s\,s^k{} \left[\int\frac{\hat{\mathrm{d}}^{2}\vect{q}\, e^{-\imath \vect{q}\cdot \vect{b}}}{[\vect{q}^2+s^2/(y^2-1)]}\right]^2\,.
\end{align}
The Fourier transform is well-known,
\begin{equation}
\int\frac{\hat{\mathrm{d}}^{2}\vect{q}\, e^{-\imath \vect{q}\cdot \vect{b}}}{\vect{q}^2+a^2}=\frac{1}{2\pi}K_0(a |b| )\,,
\end{equation}
where $K_0$ is a modified Bessel function of the second kind. It is perhaps interesting to note that this implies that the Fourier transform of $\Im \mathcal{M}(s,q^2)$ decays exponentially as $e^{-2 s |b|}$, thus rendering the integrals finite assuming $\tilde{\rho}(s)$ grows at most polynomially which can be arranged using proper subtraction terms in the KL representation.
Performing the Fourier integral, we find
\begin{align}
    	I_{k,0}={}&\frac{1}{4\pi^2(y^2-1)}	\int_{0}^\infty \mathrm{d}s\,s^k K_0\left[\frac{sb}{\sqrt{y^2-1}}\right]^2=\frac{ 
   \left(y^2-1\right)^{\frac{k-1}{2}}
   \Gamma \left(\frac{k+1}{2}\right)^3}{16\pi
   ^{3/2} \Gamma
   \left(\frac{k}{2}+1\right)}\frac{1}{|b|^{k+1}}\,.
\end{align}
Taking repeated derivatives yields the desired result
\begin{equation}
    I_{k,n}=\frac{(-1)^n 4^{n-2} 
   \Gamma \left(\frac{k+1}{2}\right)^3 
   \left(\frac{k+1}{2}\right)_n^2}{\pi ^{3/2} \Gamma
   \left(\frac{k}{2}+1\right)}\frac{\left(y^2-1\right)^{\frac{k-1}{2}}}{|b|^{2n+1+k}},
\end{equation}
where $(x)_n$ is the rising factorial.

\bibliographystyle{JHEP}
\bibliography{cite.bib}
\end{document}